\documentclass[twocolumn]{revtex4-1}
\usepackage{latexsym}
\usepackage{amsmath}
\usepackage{times}
\usepackage{amssymb}
\usepackage{fancyheadings}
\usepackage[T1]{fontenc}
\usepackage[utf8]{inputenc}
\usepackage{fancyhdr}
\usepackage{url}
\usepackage{hyperref}
\usepackage{color,longtable}
\usepackage{mydef}
\def\one{\xi}
\def\ml{-}
\def\mr{+}
\def\mleft{^{\ml}}
\def\mright{^{\mr}}
\def\ileft{_I^{\ml}}
\def\iright{_I^{\mr}}
\def\zleft{_0^{\ml}}
\def\zright{_0^{\mr}}
\newcommand{\jumpz}[1]{[#1]_0}
\newcommand{\jumpi}[1]{[#1]_I}
\newcommand{\avei}[1]{\lbrace#1\rbrace_I}
\newcommand{\avez}[1]{\lbrace#1\rbrace_0}
\def\eqdef{:=}

\usepackage{graphicx}

\makeatletter
\def\p@section{}
\def\p@subsection{}
\def\p@subsubsection{}
\makeatother

\begin{document}


\title{\bf Multiscale modeling of solar cells with
      interface phenomena} 
\author{David~H. Foster$^{1}$, Timothy Costa$^{2}$}
\author{Malgorzata Peszynska$^2$\thanks{Corresponding author. 
Email: mpesz@math.oregonstate.edu,  Tel.: +1 541 7379847, Fax: +1 519 7370517.}}
\author{Guenter Schneider$^{1}$}
\affiliation{$^1$Department of Physics, Oregon State University,
Corvallis, OR, 97331}
\affiliation{$^2$Department of Mathematics, Oregon State University,
Corvallis, OR, 97331}
\date{\today}


\begin{abstract}
\noindent 
We describe a mathematical model for heterojunctions
in semiconductors which can be used, e.g., for modeling higher
efficiency solar cells. The continuum model involves well-known
drift-diffusion equations posed away from the interface. These are
coupled with interface conditions with a nonhomogeneous jump for the
potential, and Robin-like interface conditions for carrier
transport. The interface conditions arise from approximating the
interface region by a lower-dimensional manifold. The data for the
interface conditions are calculated by a Density Functional Theory
(DFT) model over a few atomic layers comprising the interface
region. We propose a domain decomposition method (DDM) approach to decouple
the continuum model on subdomains which is implemented in every step of
the Gummel iteration. We show results for \matcigs, \matsiz,
and \matsig\ heterojunctions.

\vspace*{2ex}\noindent\textit{\bf Keywords}: Semiconductor modeling, Solar Cells, 
Materials science, Multiscale Modeling, Density Functional Theory, Drift-Diffusion Equations, Domain Decomposition, Schur Complement, Finite Differences.
\end{abstract}

\maketitle

\thispagestyle{fancy}

\section{Introduction}
\label{sec-intro}

This paper describes an interdisciplinary effort towards developing
better computational models for simulation of heterojunction
interfaces in semiconductors. The motivation comes from collaborations
of computational mathematicians and physicists with material scientists
who build such interfaces and examine their properties for the
purpose of building more efficient solar cells.  Ultimately the
design of the devices with complicated geometries and/or new
parameters must be supported by computational simulations.

Below we overview the relevant technological and physics concepts
and summarize the computational modeling challenges addressed
in this paper.

\vspace*{0.2cm} {\bf Technology background.} The experimental and
computational search for more efficient solar cells can be divided
into two approaches: a) the discovery and design of new materials on
which to base the existing thin-film solar cell designs (second
generation photovoltaics), and b) the minimization of loss mechanisms
inherent in the design of current solar cell designs through the
realization and exploitation of new physical effects (third generation
photovoltaics) \cite{Green06}.

\begin{figure}[h]
\includegraphics[width=6cm]{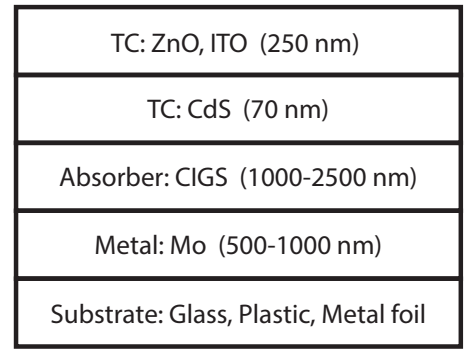}
\caption{\label{fig-CIGSstack}Schematic cross section of a
  CIGS thin-film solar cell. Light enters from the top. The top
  two layers are n-type transparent conductors (TC) consisting of
  either zinc oxide (ZnO) or indium tin oxide (ITO) and cadmium sulfide
  (CdS). Photon absorption and electron-hole generation takes place in
  the absorber layer consisting of CIGS. The bottom contact
  consists of Molybdenum metal, which sits on a fabrication dependent
  substrate material. Figure is not to scale and thickness of the
  layers is approximate. Important details such as material gradients
  in layers created through variable doping as well as alloying
  effects at some interfaces are not shown.}
\end{figure}

A thin-film solar cell is built around a semiconductor layer, where
the light is absorbed and charge carriers (electron-hole pairs) are
created, sandwiched by, on one side, the electron conducting layers,
and on the other side, by hole conducting layers. At least one of
these conducting layers must be transparent to the solar spectrum for
light to reach the central ``absorber'' layer, see
Figure~\ref{fig-CIGSstack}. Each interface between two layers in a
thin-film solar cell constitutes a {\em heterojunction}, which must be
carefully tuned to optimize the overall performance of the solar cell.
Current thin-film solar cells are built around either cadmium telluride
(CdTe) or copper indium gallium selenide, CuIn$_{1-x}$Ga$_{x}$Se$_2$,
(CIGS) as semiconductors for the absorber layer. Large scale
deployment of these established technologies is potentially impacted
by the toxicity (e.g., cadmium) or relative expense (e.g., indium) of
some of the constitutent materials.  Hence, the search for alternative
absorber semiconductors focuses on earth abundant constituents that
are nontoxic \cite{gs-scsearch_1,gs-scsearch_2}. For each new
candidate material, new conducting layers, one of which has to be
transparent, must be found and matched to each other by optimizing the
heterojunction interface between each pair of layers.

\begin{figure}[h]
\includegraphics[width=7cm]{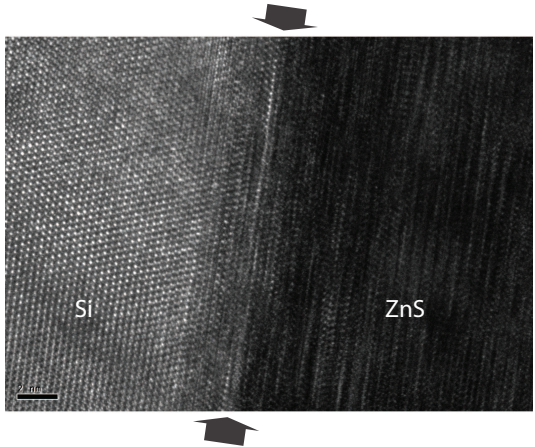}
\caption{\label{fig-TEM}Transmission electron microscope (TEM) image
  of a heterojunction between silicon (Si) and zinc sulfide (ZnS). A
  hetero-epitaxial ZnS film was grown on (111) Si via pulsed laser
  deposition. Arrows mark the interface region $\Omega^I$ between the
  two materials where the heterojunction is formed. Picture courtesy of
  Chris Reidy and Janet Tate \cite{ReidyTate}. See
  Figure~\ref{fig-domain} for idealized geometry used in the continuum
  computational model with $\Omega^I \approx I$, and
  Figure~\ref{fig-bandoffset} for schematic of microscale
phenomena in $\Omega^I$.}
\end{figure}

In current solar cell design an absorbed photon creates exactly one
charge carrier pair with an energy equal to the bandgap $E_g$ of the
absorber semiconductor material. The excess energy of the photon
$E_\mathrm{ph}-E_g$ is lost as heat. Photons with energies
$E_\mathrm{ph}\ge 2E_g$ in principle have enough energy to excite two
charge carrier pairs. Multiple charge carrier generation from high
energy photons is known as {\it impact ionization} (II) and is present
in all semiconductors but not very efficient. Exploiting quantum
effects to enhance II is an active area of research
\cite{gs-II-lit_1,gs-II-lit_2,gs-II-lit_3}. It has been hypothesized
that II can be efficient at the heterojunction interface of a low
bandgap semiconductor host material (bandgap $E_\mathrm{host}$) and a
wide bandgap semiconductor with a bandgap at least twice as big as the
bandgap of the host, i.e., $E_g > 2 E_\mathrm{host}$. We refer to this
hypothetical process as {\it heterojunction assisted impact
  ionization} (HAII) and a search for HAII is an ongoing research
effort; see Figure~\ref{fig-TEM} for a model heterojunction for HAII
consisting of the wide bandgap direct semiconductor zinc sulfide ZnS
and the low bandgap 'host' semiconductor silicon Si.

A heterojunction is characterized by several parameters that determine
its physical properties and affect the performance of a device. Most
important are the valence band and conduction band discontinuities
$\Delta E_C$ and $\Delta E_V$ (summarily referred to as band offsets,
see Table~\ref{tab-delta}), which can be obtained, in principle,
experimentally. However, computations allow for a broader search for
better photovoltaic devices.

{\bf Background on mathematical and computational models.} 
The well known \emph{drift-diffusion model} is the most widely used
continuum mathematical model for semiconductor devices, and, in particular,
for solar cells.  We refer to
\cite{Selber,markowich86,markowich,BankRoseF,jerome-sdbook,Jerome09}
for extensive background and recent extensions. It can be derived from
semi-classical transport theory based on the Boltzmann equation
together with the Poisson equation and a number of assumptions, most
importantly the introduction of a phenomelogical relaxation time and
thermal equilibrium for the charge carriers. Even though these
assumptions limit the validity of the drift-diffusion model to low
energies and longer time scales, it generally provides an adequate
description of the steady-state transport in solar cells, except in
cases where its assumptions are explicitly violated, as is generally
the case for all approaches to solar energy conversion that attempt to
harvest high energy photons more efficiently (so called third
generation photovoltaics \cite{Green06}). More sophisticated models
that includes high field and high energy effects as well as short
time-scale phenomena, include hydrodynamic transport models that go
beyond the Boltzmann transport equation (see for example Ref. 2 in
\cite{FischettiLaux88}), as well as particle based Monte
Carlo models \cite{FischettiLaux88,Saraniti06}.  See also
\cite{Gamba06} for an example of coupling and comparison of
hydrodynamic and Monte Carlo models. A computationally efficient
approach would treat only the {\emph critical} regions with a more
sophisticated model, which is coupled to the standard drift-diffusion
model to create a complete device simulation. Such an approach is the
ultimate aim of our work on heterojunctions but is outside the scope
of the present paper.

Despite the simplifying assumptions in the drift-diffusion system, it
is quite complex, and presents challenges for analysis, numerical
discretization, and nonlinear solver techniques.  The difficulties
include the nonlinear coupled nature of the system, the presence of
boundary and interior layers, and the out-of-double precision scaling
of data and unknowns, which render the model difficult to work with for
computational scientists without prior experience.

\begin{figure}[h]
\includegraphics[width=8cm]{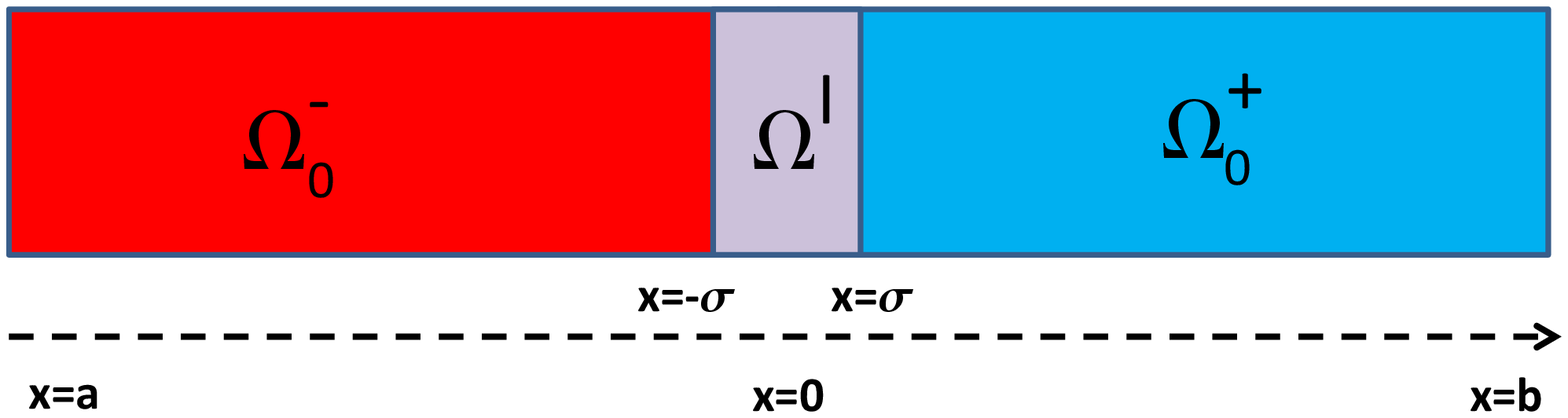}
\\
\includegraphics[width=8cm]{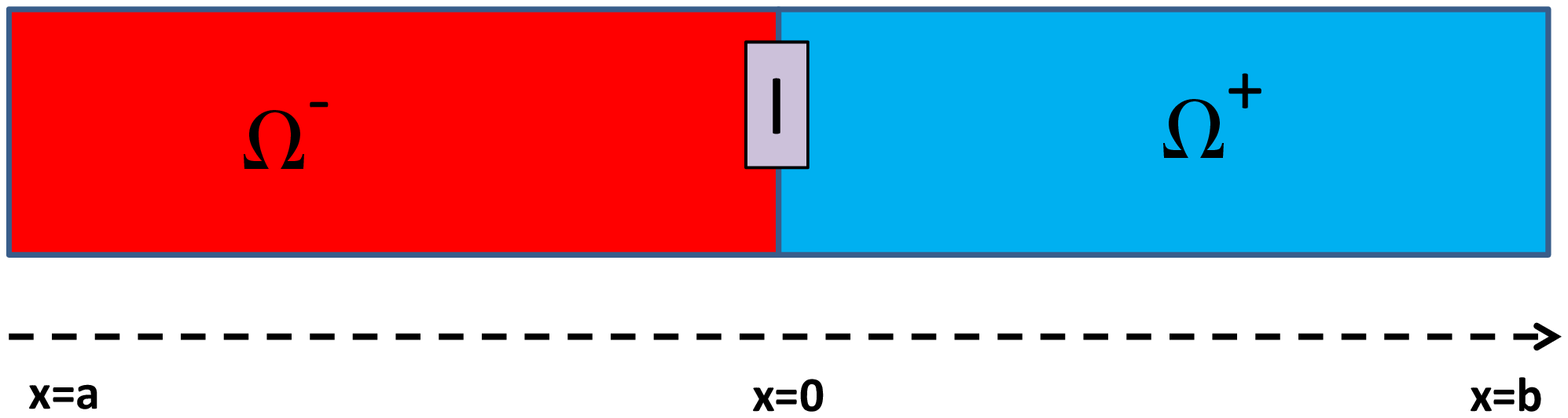}
\caption{\label{fig-domain}Geometry of heterojunction interface domain
  $\Omega=(a,b)$. {\bf Top:} $\Omega_0\mleft=(a,-\sigma)$ is made
  primarily of atoms of first material, and
  $\Omega_0\mright=(\sigma,b)$ corresponds to the second material. The
  crystalline structure in the interface region
  $\Omega^I=(-\sigma,\sigma)$ is uneven and the continuum model does
  not hold across $\Omega^I$, but its width $2\sigma\ll b\!-\!a$ is
  small compared to that of $\Omega$.  {\bf Bottom:} It is convenient
  to consider an approximation $I \eqdef\{0\} \approx \Omega^I$ with
  $\Omega\mleft=(a,0)$ and $\Omega\mright=(0,b)$.  See
  Figure~\ref{fig-TEM} for an example of a physical heterojunction in
  \matsiz, Figure~\ref{fig-bandoffset} for schematic, and
  Figure~\ref{fig-SiZnS} for DFT and continuum model results for
  \matsiz\ illustrating phenomena in $\Omega^I$ and $\Omega$,
  respectively. }
\end{figure}

The presence of a heterojunction adds to that complexity. Consider a
1d semiconductor region $\Omega=(a,b)$ made of two materials, with an
{\em interface region} $\Omega^I=(-\sigma,\sigma)$ located near $x=0$,
as shown in Figure~\ref{fig-domain}.  Processes in $\Omega^I$ where
the two materials meet are characterized, e.g., by steep gradients and
discontinuities of the primary variables, and cannot be resolved on
the scale of drift-diffusion models.

One way to model heterojunctions is via an atomic scale model such
as Density Functional Theory (DFT) in $\Omega^I$ which however cannot
simulate more than a few atomic layers
\cite{MartinVanDeWalle1}. Alternatively, one can use an
approximation of the interface region by a lower-dimensional interface
$I$ \cite{Yang,HorioYanai}, as in Figure~\ref{fig-domain}, along with
a separate mathematical model approximating the physical phenomena
across the interface.  This is described, e.g., in
\cite{HorioYanai,Yang}, and heterojunction models have been
implemented, e.g., in community codes such as 1D semiconductor
modeling programs AMPS \cite{ampsweb} and SCAPS
\cite{Berg2000,scapsweb}. However, literature and documentary material
for these as well as early modeling references such as
\cite{HorioYanai,Yang,fonash1979} do not analyze the mathematical
assumptions underlying the treatment of the interface, and many are
quite subtle and unusual.
 
Ideally, one would find a way to tightly couple the continuum model
away from the interface, i.e., in $\Omega_0\mleft \cup \Omega_0\mright$
with {\em some other} model in the interface region $\Omega^I$, but this is
not feasible yet. Instead, in this paper we take a step in the
direction of a future coupled continuum--discrete model by decomposing
functionally the continuum heterojunction model into subdomain parts
on $\Omega\mleft$, $\Omega\mright$ and the interface part $I$. In
simulations we use realistic interface parameters computed by the DFT
model on $\Omega^I$ which is, however, entirely decoupled from
the continuum model.

A substantial part of this paper is devoted to the careful modeling of
the interface equations elucidating the challenges and unusual
features as compared to the traditional transmission conditions in
which the primary variables and their normal fluxes are continuous.
We also reformulate the heterojunction model using domain
decomposition method (DDM) \cite{quarteroniV} which, to our knowledge,
has not been applied to heterojunction models. (Throughout the paper
we use DDM to denote concepts related to domain decomposition, in an
effort to avoid confusion with the drift-diffusion equations).  We
present preliminary results of our DDM algorithm as applied to each step
of the Gummel loop for homojunctions as well as to the potential
equation for heterojunctions, and these results are promising.

DDM requires that we carefully examine the behavior of the primary
variables and their fluxes across the interface. In fact, the former
lack continuity, and the primary variables either have a step jump, or
satisfy a nonlinear Robin-like condition. We find similarities of the
heterojunction model to various fluid flow models that have recently
attracted substantial attention in the mathematical and numerical
community. Analysis in \cite{MoralesS10, MoralesS12} and modeling and
simulations for flow across cracks and barriers in
\cite{Roberts05,RobertsFrih} have been pursued. See also recent
numerical analysis work in
\cite{GiraultRW05,GiraultR09,RiviereK10,RiviereA12} devoted to other
flow interface problems. Still, the heterojunction problem appears
even more complex than (some of) those listed above due to its coupled
nature and nonlinear form of the interface conditions as well as 
to the complexities of the subdomain problems. 

Clearly nontrivial mathematical and computational analyses following
\cite{MoralesS10,MoralesS12,Roberts05,RobertsFrih} as well as
semiconductor-specific implementations are needed, but the formulation
given here opens avenues towards applications of modern numerical
analysis techniques beyond the finite differences that have been
traditionally employed. 
In particular, the DDM solver given here can be easily extended to multiple
interfaces or complex 2d geometries.  In contrast, such extensions may
be very difficult for monolithic solvers in which the interface equations
are hard-coded. We plan to address 2d implementation in our future work.

The outline of the paper is as follows. We present an overview of the
DFT model in Section~\ref{sec-dft}. Detailed description of the
subdomain and heterojunction interface models is given in
Section~\ref{sec-model}. Here we also describe the domain decomposed
formulation of the problem in which the interface problem is isolated
in its own algebraic form amenable to an iterative solver, with
particular care paid to the heterojunction formulations. In
Section~\ref{sec-results} we present computational results for the
DFT, monolithic, and DDM solvers. In Section~\ref{sec-conclusions} we
discuss and summarize the results. The Appendix in
Section~\ref{sec-appendix} contains some auxiliary calculations and
data which support the developments in Sections~\ref{sec-model} and
\ref{sec-results}.

\vspace*{0.2cm}
\section{Computational Model: DFT}
\label{sec-dft}

Quantum mechanics of the electrons governs the properties of matter
and hence the properties of a heterojunction. The direct solution of
the quantum mechanical problem of an interacting $N$ electron system
remains an intractable problem, but the reformulation in terms of the
electron density of Hohenberg and Kohn and Kohn and Sham \cite{HK,KS}
provides an indirect and, using appropriate approximations, feasible
approach for many problems in condensed matter theory, materials
science, and quantum chemistry. Density functional theory (DFT) has
become the standard approach to calculate material
properties from first principles. To set the stage for the calculation
of heterojunction parameters and in particular band offset energies,
we give a condensed overview of DFT loosely following and adopting the
notation of \cite{PerdewKurthDFTprimerbook}. For details we refer the reader
to one of several monographs and reviews \cite{DFTprimerbook,
DFTadvancedbook}. 

\subsection{Density Functional Theory} 
\label{sec-dfttheory}
The large mass difference between electrons and nuclei
allows us to treat the motion of the light electrons relative to a
background of nuclei with fixed positions. DFT deals with the standard
Hamiltonian of $N$ interacting electrons 
(ignoring spin for brevity)
\ba
\label{eq-dft-hamiltonian}
\hat{H} = \hat{T} + \hat{V}_{ee} + \hat{V}_\mathrm{ext},
\ea
which consists of the kinetic energy operator $\hat{T}$, 
\bas
\label{eq-kinetic-energy}
\hat{T} = \sum_{i=1}^{N} -\frac{\hbar^2}{2m}\nabla_i^2,
\eas
the Coulomb interaction between the electrons,
\bas
\label{eq-ee-coulomb}
\hat{V}_{ee} = \sum_{\substack{i,j=1\\i<j}}^{N} 
                      \frac{e^2}{|{\bf \hat{r}}_i - {\bf \hat{r}}_j|} .
\eas
and the interaction of the electrons with an external potential
\bas
\label{eq-Vext}
\hat{V}_\mathrm{ext} = \int d^3 r v_\mathrm{ext}({\bf r})
\hat{n}({\bf r}); \;\;
\hat{n}({\bf r}) = \sum_{i}^{N} \delta({\bf r} - {\bf \hat{r}}_i).
\eas
The external potential $v_\mathrm{ext}({\bf r})$ contains the
contributions from the atomic nuclei and possibly other terms. 
Here $\hbar = h/2\pi$, where $h$ is the Planck constant, $m$ is the electron
mass, $-e=-|e|$ is the charge of the electron, and $\hat{\mathbf{r}}_j $ is
the quantum mechanical position operator for electron $j$
\cite{messiah99book}.

The solutions of the stationary Schr\"odinger equation,
\ba
\label{eq-schroedinger}
\hat{H} |\Psi_n\rangle = E_n |\Psi_n\rangle,
\ea
are the many-electron wavefunctions $|\Psi_n\rangle,$ with energy $E_n$. For the
ground state $|\Psi_0\rangle,$ with energy $E_0$, the Schr\"odinger
equation \eqref{eq-schroedinger} is equivalent to a variational
principle over all permissible  $N$ electron wavefunctions,
\ba
\label{eq-dvpfull}
E = \underset{\Psi}{\mathrm{min}} \langle \Psi|\hat{H}|\Psi \rangle .
\ea
The theorem by Hohenberg and Kohn\cite{HK,LL_1,LL_2} provides the existence of a
density variational principle for the ground state, 
\ba
\label{eq-dvp}
E = \underset{n}{\mathrm{min}}
\left\{ 
F[n] + \int d^3 r v_\mathrm{ext}({\bf r}) n({\bf r}) 
\right\},
\ea
where
\ba
\label{eq-ffunctional}
F[n] = \underset{\Psi\rightarrow n}{\mathrm{min}} 
           \langle \Psi|\hat{T}+\hat{V}_{ee}|\Psi \rangle ,
\ea
is a universal functional defined for all $N$ electron densities $n({\bf r})$.
The ground state density (or densities if the ground state is
degenerate) that minimizes \eqref{eq-dvp}, uniquely defines
the external potential $v_\mathrm{ext}({\bf r})$. 
Expressing ground state properties in terms of the electron density 
as in \eqref{eq-dvp}
defined over $\R^3$ instead of a fully antisymmetric $N$ electron
wavefunction defined over $\mathbb{C}^{3N}$ as in \eqref{eq-dvpfull}
is a huge reduction in complexity,
but this simplification comes at a cost, since 
the functional $F[n]$ is not known.

DFT requires suitable approximations for $F[n]$ as well as efficient
minimization schemes to determine the approximate ground state energy
and electron density. The Kohn-Sham equations~\cite{KS} described next
provide an iterative solution to the minimization problem and serve as
the basis for approximate density functionals.

\subsection{Kohn-Sham equations} 
\label{sec-kseqn}
A solution to \eqref{eq-dvp} can be found for a system of {$N$}
non-interacting electrons governed by effective single electron
Schr\"odinger equations,
\begin{multline}
\label{eq-ks-eqn}
\left(
-\frac{\hbar^2}{2m}\nabla^2 +
v_\mathrm{es}({\bf r}) +
v_\mathrm{xc}([n];{\bf r})
\right) 
\psi_k({\bf r}) \\= \varepsilon_k \psi_k({\bf r}),
\end{multline}
where $\psi_k$, $\varepsilon_k$ denote eigenstate and energy
of a single particle.
Here we have introduced the electrostatic potential
\bas
\label{eq-electrostatic-potential}
v_\mathrm{es}({\bf r}) =
\int d^3 r' \frac{n({\bf r}')}{|{\bf r}-{\bf r}'|} +
v_\mathrm{ext}({\bf r}) ,
\eas
consisting of the Hartree term of the Coulomb interaction between the
electrons and the external potential.
The electron density is given by
$
n({\bf r}) = \sum_k \theta(E_F-\varepsilon_k) |\psi_k({\bf r})|^2,
$
and the Fermi energy $E_F$ is determined by the normalization condition
for the density
$N= \int d^3 r n({\bf r})$.
Approximations for the functional $F[n]$ from \eqref{eq-ffunctional}
enter through the
exchange-correlation potential $v_{xc}({\bf r})$, which is the
functional derivative of the exchange correlation energy
$E_\mathrm{xc}[n]$ with respect to the electron density:
\ba
\label{eq-xc-potential}
v_\mathrm{xc}([n];{\bf r}) = 
     \frac{\delta E_\mathrm{xc}[n]}{\delta n({\bf r})} .
\ea
The exchange correlation energy $E_\mathrm{xc}[n]$ is the remainder
of the functional $F[n]$,
after the kinetic energy of the $N$ non-interacting electrons
\begin{multline*}
T_\mathrm{ni}[n] = \sum_k \theta(E_F-\varepsilon_k) \int d^3 r \psi_k^{*}({\bf r})
\left(-\frac{\hbar^2}{2m}\nabla^2\right) \psi_k({\bf r}) ,
\end{multline*}
and the Hartree energy
\bas
\label{eq-hartree}
U[n] = \frac{1}{2} \int d^3 r \int d^3 r' 
           \frac{n({\bf r}) n({\bf r}')}{|{\bf r}-{\bf r}'|} ,
\eas
have been subtracted:
\ba
\label{eq-xc-energy}
E_\mathrm{xc}[n] = F[n] - T_\mathrm{ni}[n] - U[n] .
\ea
A self--consistent solution of the Kohn-Sham equations
\eqref{eq-ks-eqn}-\eqref{eq-xc-energy} can be found iteratively for a
suitable choice of approximation of the exchange correlation energy
$E_\mathrm{xc}[n]$. 

A particularly simple and surprisingly effective
approximation is the local density approximation (LDA)
\ba
\label{eq-lda}
E_\mathrm{xc}^\mathrm{LDA}[n] = \int d^3 r \, n({\bf r})
e_\mathrm{xc}(n({\bf r})) ,
\ea
where $e_\mathrm{xc}(n({\bf r}))$ is the exchange correlation energy of
the uniform electron gas \cite{electrongas_1,electrongas_2}.

Finally, the total ground state energy of the $N$ electron problem
given by \eqref{eq-dft-hamiltonian}-\eqref{eq-ffunctional}
can be calculated as
\ba
\label{eq-ks-total-energy}
E = T_\mathrm{ni}[n] +\!\int\!d^3 r n({\bf r}) v_\mathrm{ext}({\bf r}) +
 U[n] + E_\mathrm{xc}[n].
\ea

The Kohn-Sham equations are a system of coupled single particle wave
equations, but their computational cost increases rapidly when
non-local approximations for the exchange-correlation potential are
used.  The Kohn-Sham equations not only provide an efficient
numerical tool to solve the density variation principle
\eqref{eq-dvp}, but they are equivalent to the well known
bandstructure equations of the independent electron approximation. It
is worthwhile to note however that DFT gives no additional
justification for the use of the independent electron
approximation. The single particle eigenstates $\psi_k$ with single
particle energies $\varepsilon_k$ appear merely as reference system to
solve the complicated interacting problem in an approximate but highly
useful way. They are used extensively because they provide excellent
results for many properties for an exceedingly large number of systems
in condensed matter physics, materials science, and quantum chemistry.

\subsection{Semiconductors properties}
\begin{figure}[h]
\includegraphics[width=6cm]{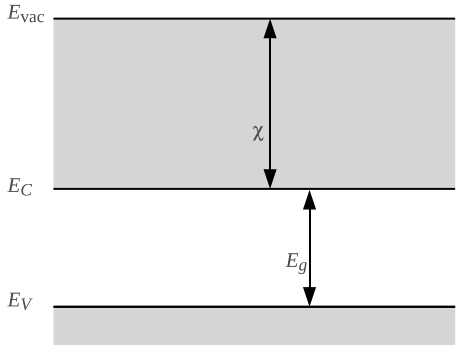}
\caption{\label{fig-banddiagram} Schematic band diagram of a
  semiconductor indicating the vaccum energy $E_{vac}$, the electron
  affinity $\chi$, the bandgap $E_g$, the conduction band energy $E_C$,
and the valence band energy $E_V$.}
\end{figure}

For an intrinsic semiconductor (undoped, zero temperature) the valence
band energy $E_V$ is the energy of the occupied single 
electron state with the highest energy, and the
conduction band energy $E_C$ is the energy of the unoccupied single-electron
state with the lowest energy. The bandgap $E_g$ is the minimum energy
required to remove one electron from the occupied single particle
states, the valence band, and add it to the unoccupied states, the
conduction band, and is given by the difference $E_g = E_C-E_V$ (see
Figure \ref{fig-banddiagram}).
The electron affinity $\chi$ denotes the energy required to add one
electron to a semiconductor and is given by the difference of the electron
vacuum energy $E_{vac}$ and the conduction band energy $E_C$. 
As is customary in semiconductor device modeling, we define the
potential $\psi \eqdef -E_{vac}$. In terms of the potential $\psi$ the
semiconductor quantities that enter the continuum model are defined as 
\ba
\label{eq-vac}
E_C\eqdef -\psi -\chi,\;
E_V \eqdef -\psi -\chi - E_g.
\ea

The bandgap of a semiconductor is a ground state property and can be
expressed in terms of ground state energies as in \eqref{eq-dvp}.  For
a semiconductor with $N$ electrons it is denoted by $E(N)$ so that
\begin{multline*}
E_g = \mathrm{min} \left\{ [E(N+1) - E(N)]- [E(N)-E(N-1)]\right\}.
\end{multline*}
In practice however, the bandgap is calculated in terms of the single
electron states and energies that are the self-consistent solutions of the Kohn-Sham
equations \eqref{eq-ks-eqn}, and
\bas
E_g = E_C - E_V = \varepsilon_\mathrm{unocc}^\mathrm{min} - 
                             \varepsilon_\mathrm{occ}^\mathrm{max}.
\eas
Local approximations of the exchange-correlation energy
\eqref{eq-xc-energy}, like the LDA \eqref{eq-lda}, typically result in
bandgaps $E_g$ that are too small \cite{eg2small}. This negative result is well
understood, and improved approximations allow the accurate prediction
of bandgaps from first principles \cite{Kresse}.

In preparation for the calculation of band offsets from periodic
supercells described in the next section, we define and calculate a
local reference energy $E_\mathrm{ref}=V_\mathrm{es}^\mathrm{avg}$,
where $V_\mathrm{es}({\bf r})$ is the electrostatic potential defined
in equation \eqref{eq-electrostatic-potential}. The electrostatic
potential is spatially averaged over the entire unitcell of the
semiconductor to determine the average electrostatic potential,
$V_\mathrm{es}^\mathrm{avg}$. Our choice of reference energy is not
unique but a convenient one for electronic structure methods based on
planewave basis sets.

\subsection{Band offset calculation} 
\begin{figure}[h]
\includegraphics[width=8cm]{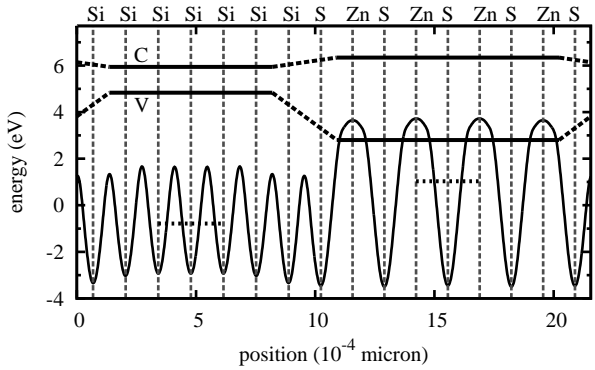}
\caption{\label{fig-bandoffset} 
Schematic cross section through the supercell of a band offset
calculation for a \matsiz\ heterojunction, with 
the interface normal in the (100) crystallographic direction. The cross section is
perpendicular to the interface and the location of individual atoms
is indicated. Periodic boundary
conditions result in 2 heterojunctions located between the Si-S layers
approximately located at 0 and $10\times10^{-4} \mu$m, which are
identical for this interface.
The planar averaged electrostatic potential is shown as full
curve, and the potential average, $V_\mathrm{es}^\mathrm{avg}$, in both
Si and ZnS is marked by horizontal dotted lines. The length of the
horizontal dotted lines indicates the central region in each semiconductor over which the
average was performed. The valence band (conduction band) energies
$E_V$ ($E_C$) for this heterojunction are indicated by solid straight
lines and have been calculated as described in the text.
A realistic supercell for a band offset calculation would be 3-4 times
longer than the one depicted here. }
\end{figure}

We formulate the heterojunction problem in terms of a supercell
surrounding a section of the interface
\cite{MartinVanDeWalle1,MartinVanDeWalle2}.  Periodic boundary
conditions are used in all spatial dimensions. The supercell
dimensions in the plane of the interface are determined by the
periodicity of both semiconductor lattices. The length of the
supercell must be sufficiently large so that away from the interface
both semiconductors have essentially bulk-like properties.  The use of
periodic boundary conditions allows the application of the same
computational tools to the interface problem that have been developed
for bulk materials. On the other hand, some complications may result from
the use of periodic boundary conditions, e.g., one ends up with two
heterojunctions which are symmetric only in simple cases such as 
for the \matsiz\ heterojunction shown in Figure
\ref{fig-bandoffset}.  
Periodic boundary conditions may be used, even in the case of two
different (asymmetric) interfaces, provided the supercell is long
enough to distinguish the rapid band behavior at a single interface
from slow, artificial changes that may occur due to the boundary
conditions.  (The slopes of the bands present in continuum model
solutions are not modeled in this calculation, and are considered
infinitesimal on the scale of the supercell.)
The interface atomic structure of the
heterojunction is generally not known and can be determined, in
principle, within the constraint of the boundary conditions of the
supercell by minimizing the energy of the supercell as a function of
the atomic positions.

To determine the valence band energy separately for both
semiconductors in the supercell, we calculate the local reference energy
$E_\mathrm{ref}=V_\mathrm{es}^\mathrm{avg}$ defined in the previous
section, but now the average over the electrostatic potential
energy is taken for each semiconductor separately over a finite region
in the supercell, where each semiconductor has essentially bulk-like
properties (see Figure \ref{fig-bandoffset}). The accuracy of the averaging procedure
can be systematically controlled by increasing the size of the
supercell used for the interface calculation. We can
relate the energy difference $E_V - E_\mathrm{ref}$ for both 
semiconductors in the supercell with the corresponding energy difference in the
separate bulk calculations. For the heterojunction shown in Figure
\ref{fig-bandoffset} we obtain, for silicon,
\bas
\label{eq-vbenergy-calc}
E_V(\mathrm{Si}) &=& E_V(\mathrm{Si}_\mathrm{bulk}) - 
                                    E_\mathrm{ref}(\mathrm{Si}_\mathrm{bulk}) + 
                                    E_\mathrm{ref}(\mathrm{Si}), 
\\
E_C(\mathrm{Si}) &=& E_V(\mathrm{Si}) + E_g(\mathrm{Si}).
\eas
Similar equations follow for the other semiconductor (ZnS in our
example). The valence and conduction band energy offsets are the
differences  
$\Delta E_V = E_V(\mathrm{Si}) - E_V(\mathrm{ZnS})$ and 
$\Delta E_C = E_C(\mathrm{Si}) - E_C(\mathrm{ZnS})$.

\label{sec-bandoffset}

\section{Computational Model: continuum}
\label{sec-model}
Here we present the continuum model for a semiconductor device with an
interface. Our presentation of the drift-diffusion models is based on
\cite{Selber,markowich86,markowich,MarkRS}, with analysis as presented in
\cite{markowich,jerome-sdbook} while the interface physics and model
have been described in \cite{Sze,HorioYanai,Yang}.

We use the geometrical representation presented in
Figure~\ref{fig-domain}. Recall that, while the approximation
$\Omega^I \approx I$ is convenient for a continuum model, it is a
simplification of the real physical situation, in which the interface
region is composed of a few atomic layers.

First we describe the drift-diffusion equations, a coupled
system of nonlinear PDEs, with coefficients which depend on the
material from which $\Omega\mleft, \Omega\mright$ are made. Next we
describe the interface model, its numerical approximation, and the
domain decomposition formulation.  The various coefficients are given
in Tables~\ref{tab-param}, \ref{tab-delta} and depend on the material
and the type of interface.

\vspace*{0.2cm}
 \begin{table}[htbp]
  \centering
  \caption{Data: material constants. See Table~\ref{tab-matprops}
for the values used in simulations.}
  \label{tab-param}
  \begin{tabular}{lcc}
    \hline
symbol    & parameter\\
    \hline
    $N_T(x)$   & net doping profile &  \\
    $\tau_n(x)$ & trap-related electron lifetime & \\ 
    $\tau_p(x)$ & trap-related hole lifetime & \\
    $R_{dc}$ & direct recombination coefficient & \\
    $\eps$   & dielectric constant \\
    $\chi$   & electron affinity \\
    $N_i$   & intrinsic carrier concentration\\
    $E_g$   & bandgap\\
    $N_C$   & density of states, conduction band \\
    $N_V$   & density of states, valence band \\
    $D_n$   & electron diffusivity\\
    $D_p$   & hole diffusivity\\
    $A_n$   & effective Richardson's constant for electrons\\
    $A_p$   & effective Richardson's constant for holes\\
    \hline
  \end{tabular}
\end{table}

\vspace*{0.2cm}
 \begin{table}[htbp]
  \centering
  \caption{Data: interface parameters.}
  \label{tab-delta}
  \begin{tabular}{lcc}
    \hline
symbol    & parameter\\
    \hline
    $\mydel \psi$ & jump of potential & \\
    $\mydel E_C = -(\mydel \psi + \mydel \chi)$ & jump of conduction band energy & \\
    $\mydel E_V = \mydel E_C - \mydel E_g$ & jump of valence band energy & \\
    \hline
  \end{tabular}
\end{table}

{\bf Notation:} We adopt the following notation for the continuum model.
The dependence on some independent or dependent variables is omitted if
it is clear from the context. In particular, we use $R_T$ for the
recombination terms $R_T=R_T(n,p)$ which depend on the variables $n$,
$p$, as well as on several position dependent parameters. The same
concerns various material parameters, whose dependence on the position
$x \in \Omega$, and in particular on the type of material, i.e.,
whether $x \in \Omega \mleft$, or $\x \in \Omega \mright$, is
dropped. When relevant, we denote material dependent constants using
superscripts $\mleft,\mright$. Also, we denote by
$\one\mleft(x)$ the characteristic function of $\Omega\mleft$, i.e.,
the function equal to one in $\Omega\mleft$ and to zero elsewhere, and
$\one\mright$ is defined analogously. These help, e.g., to write a
piecewise constant material dependent coefficient, e.g.,
$\eps(x)=\eps\mleft\one\mleft(x) +\eps\mright\one\mright(x)$.

We use notation $\nu$ to denote a unit vector normal to a boundary or
interface pointing {\it outward} to the given domain.

We distinguish the value $s\ileft$ of a physical quantity $s$
evaluated on the left side of $I$. With the geometry
as defined above we have 
\ba
\label{eq-i}
s\ileft \eqdef \lim_{x\to -\sigma\mleft}s(x),\;
s\iright \eqdef \lim_{x\to \sigma\mright}s(x). 
\ea
We also use the notation 
\bas
\jumpi{s} \eqdef s\iright-s\ileft,\;\;
\avei{s} \eqdef \frac{s\iright+s\ileft}{2}. 
\eas

Similar notation, when $\sigma=0$, is common in computational
mathematics and in particular in numerical analysis of Discontinuous
Galerkin (DG) Finite Element methods \cite{riviere-book}, where the symbols
\bas
\jumpz{s}& \eqdef& s(0\mright)-s(0\mleft)\eqdef \lim_{x\to 0\mright}s(x)
- \lim_{x\to 0\mleft}s(x),\;
\\
\avez{s} &\eqdef& \frac{s\zright+s\zleft}{2},
\eas
are used at computational nodes (here at $x=0$).  The context in which
we use $\jumpi{s},\avei{s}$, is different from the DG-specific use of
$\jumpz{s},\avez{s}$, respectively, since $\sigma \neq 0$. 

To make this distinction clear, in the model derivations we use
$\jumpz{\cdot}$ across a {\em homojunction} interface $I$, and
$\jumpi{\cdot}$ across a {\em heterojunction} $I$ approximating some
$\Omega^I$.  The case of a {homojunction} and $\sigma=0$ is when
the materials in $\Omega\mleft, \Omega\mright$ are the same, but the
doping characteristics $N_T$ change drastically across $I$. The case
of a heterojunction and $\sigma \neq 0$ is when the materials are
different, and there are additional physical phenomena that need to be
accounted for in $\Omega^I$, i.e., across $I$.

Throughout the paper we use, to the extent possible, nondimensional
quantities, while keeping material-dependence evident through
notation. Our use of nondimensional quantities is consistent with
typical scaling applied in semiconductor modeling such as described in
\cite{Selber}. When needed for clarity, we emphasize this by
referring to the ``scaled units''.

Finally, we use notation of functional spaces as traditionally
adopted, e.g., in \cite{ShowDover,BrezziFortin,quarteroniV}. In
particular, $C^k(\Omega)$ is a space of functions of up to $k$
continuous derivatives on $\Omega$.  For weak formulations we use
Sobolev spaces $H^1(\Omega) \eqdef \{ w: w, \nabla w \in L^2(\Omega)
\}$ instead of $C^2(\Omega)$, where we recall $L^2(\Omega) \eqdef
\{ w: \int_{\Omega} w^2 dx < \infty\}$. Also, $L^{\infty}(\Omega)$ is
the space of essentially bounded functions.

\subsection{Bulk equations in a homogeneous semiconductor}

We assume isothermal and steady-state regimes.  While transient
behavior has decayed, the time-independent transport of electrons and
holes is described by the spatially-dependent electron and hole
currents.  These currents are steady-state responses to certain
boundary conditions (applied voltages), bulk carrier generation due to
illumination, and other carrier sources and sinks such as
electron-hole recombination.

For convenience we are presenting the continuum model in terms of
dimensionless quantities.  Each quantity is scaled by a dimensioned
quantity and $\bold x$ may be scaled by a length quantity. The scaling
is discussed further in Appendix \ref{sec-appendix-scale}.

The drift-diffusion model in a single-material semiconductor domain is
\ba
\label{eq-psi}
- \nabla \cdot (\eps \nabla \psi) &= &q\eqdef p-n+N_T,
\\
\label{eq-ncont}
\nabla \cdot J_n &=& R_T, 
\\
\label{eq-pcont}
\nabla \cdot J_p &=& -R_T. 
\ea
Here $\psi,p,n$ are, respectively, the potential and the charge
densities of holes and electrons, and $N_T(x)= N_D(x)-N_A(x)$ is the
given net doping profile including the donor $N_D$ and acceptor doping
$N_A$. The recombination term $R_T\eqdef R_T(n,p)$ is defined below.

The current density $J_n = J_{n,drift} + J_{n,diff}$ 
where the drift part is due to the electric displacement field
$ J_{n,drift}=-D_n n \nabla \psi $
and the diffusive part
$ J_{n,diff}=D_n \nabla n$. Thus we have
\ba 
\label{eq-jn}
J_n = D_n (-n \nabla \psi + \nabla n).
\ea 
Similarly, the flow of holes is described by
\ba 
\label{eq-jp}
J_p = -D_p( p \nabla \psi + \nabla p).
\ea 

For convenience of numerical computations the currents can be
defined with the use of the quasi Fermi potentials $\psi_n,\psi_p$ as
\ba
\label{eq-psin}
J_n &= - D_n n \nabla \psi_n,
\\
\label{eq-psip}
J_p &= - D_p p \nabla \psi_p,
\ea
where $n,\psi_n$ and $p,\psi_p$ are related via Maxwell-Boltzmann statistics
\ba
\label{eq-npsin}
n &=& N_C \exp(\psi + \chi -\psi_n),
\\
\label{eq-ppsip}
p &=& N_V \exp(-\psi - \chi +\psi_p-E_g).
\ea

To see why \eqref{eq-psin} and \eqref{eq-jn} are equivalent, we
differentiate \eqref{eq-npsin} to see $\nabla n = n
\nabla(\psi-\psi_n)$.  Of course, this change of variable is only
possible if $n, \psi, \psi_n$ are differentiable, and, in particular,
is not true at heterojunctions {\it across} $I$.

The recombination term $R_T(n,p)$ is given as in \mpcite{Sze}{Sec 1.5.4} by
\ba R_T = - G + R_d + R_{SRH}.
\ea 
Here $G=G(x)$ is a position-dependent carrier generation source term
from the light sources. The terms $R_d$ and $R_{SRH}$ are given traditionally as
Shockley-Reed-Hall recombination terms
\ba
R_d &= R_{dc}(np - N_i^2),
\ea
\ba
R_{SRH}&=\frac{np-N_i^2}{\tau_p(n+N_i) + \tau_n(p+N_i)}.
\ea
The parameters $\tau_p,\tau_n$ are material constants given in
Table~\ref{tab-param}, and $N_i^2 \eqdef N_CN_V \exp(-E_g)$.

When analyzing well-posedness, or numerically solving the model
\eqref{eq-psi}--\eqref{eq-pcont}, one has to make a decision on the
choice of primary variables. While $\psi,n,p$ are most physically
natural, two other sets of variables $\psi,\psi_n,\psi_p$, as well as
so-called Slotboom variables $\psi,u,v$ can be used.  The Slotboom
variables $(u,v)$ are defined as
\ba
\label{eq-slotboom}
n = \delta^2 \exp(\psi) u,\;
p = \delta^2 \exp(-\psi) v,
\ea
where $\delta$ is a scaling parameter that depends on the material
as well as the doping profile $N(x)$.  Similarly to
\eqref{eq-npsin}--\eqref{eq-ppsip}, this change of variables only
works if $\psi,u,v$ are smooth, thus, not across $I$.

Finally, we define the effective Fermi levels $E_{Fn}\eqdef -\psi_n$,
$E_{Fp} \eqdef - \psi_p$, where appropriate scaling to eV units is
implicit.  We further obtain from \eqref{eq-npsin}, \eqref{eq-ppsip}
\ba
\label{eq-nEFn}
n &= N_C \exp (E_{Fn} - E_C),  
\\
\label{eq-pEFp}
p &= N_V \exp (E_V - E_{Fp}).
\ea
It is also convenient to define, via \eqref{eq-psin}
\ba
\label{eq-dndp}
D^n \eqdef D_n n; D^p \eqdef D_p p,
\ea
which by \eqref{eq-npsin} depend nonlinearly on $\psi_n,\psi_p$, respectively.

\subsection{External boundary conditions}
\label{sec-bc}

To complete the model \eqref{eq-psi}-\eqref{eq-pcont} as a boundary
value problem, we need external boundary conditions on $\partial
\Omega= \{a,b\}$. What follows is a summary of, e.g.,
\mpcite{markowich86}{Sec. 2.3}. 

We use Dirichlet conditions for \eqref{eq-psi},
\ba 
\label{eq-bc0}
\psi\vert_{\partial \Omega}=\psi_D; \; \psi_D(a)=\psi_a, \; \psi_D(b)=\psi_b.
\ea
To determine physically meaningful values of $\psi_a,\psi_b$ one finds
first the neutral-charge thermal equilibrium values of $\psi_{a}^{TE},
\psi_{b}^{TE}$, i.e., solving, e.g, at $x=a$, the algebraic problem
solved for $\psi_a^{TE}$
\begin{multline*}
N_V \exp(-\psi_{a}^{TE} - \chi - E_g) 
\\- N_C \exp(\psi_a^{TE} + \chi) + N_T\vert_{x=a} = 0.
\end{multline*}
This corresponds to setting $\psi_n = \psi_p = 0$ everywhere on
$\Omega$, and dropping the derivatives from \eqref{eq-psi}. At $x=b$
we solve a similar equation for $\psi_b^{TE}$. The neutral-charge
thermal equilbrium conditions are appropriate for sufficiently long
single material domains with ``ideal'' contacts with external metal
regions.

With $\psi_a^{TE},\psi_b^{TE}$ we set 
\bas
\psi_a=\psi_a^{TE} + V_a,\;
\psi_b=\psi_b^{TE} + V_b, 
\eas
where $V_a$ and $V_b$ are physically controllable external (scaled)
voltages; see Section~\ref{sec-dfresults} for their use.

The boundary conditions for \eqref{eq-ncont}-\eqref{eq-pcont} are
specified using the individual carrier currents via contact-specific
effective recombination velocities $v_{n,a}$, $v_{p,a}$, $v_{n,b}$,
and $v_{p,b}$.  In scaled units these Robin conditions read
\ba
\label{eq-bc1}
J_n \cdot \nu\vert_{x=a} &=  -v_{n,a} (n- n_0)\vert_{x=a}, 
\\
\label{eq-bc2}
J_p \cdot \nu\vert_{x=a} &=  v_{p,a} (p - p_0)\vert_{x=a}, 
\\
\label{eq-bc3}
J_n \cdot \nu\vert_{x=b}&=  -v_{n,b} (n - n_0) \vert_{x=b}, 
\\
\label{eq-bc4}
J_p \cdot \nu\vert_{x=b} &=  v_{p,b} (p - p_0)\vert_{x=b}.
\ea
Here $n_0,p_0$ are the carrier densities corresponding to the thermal
equilibrium values $\psi_n=\psi_p=0$ via \eqref{eq-npsin}-\eqref{eq-ppsip}.

\subsection{Well-posedness in a single material}
\label{sec-analysis}

We recall now after \cite{markowich} the basic information concerning
well-posedness of the system. The Gummel iteration introduced here is
relevant for the numerical solver as well as interface
decomposition procedure.

Let $H \colon = H^1(\Omega) \cap L^\infty(\Omega)$, with the norm
inherited from $H^1(\Omega)$. To analyze existence and uniqueness of
solutions to \eqref{eq-psi}--\eqref{eq-pcont}, under boundary
conditions \eqref{eq-bc0}-\eqref{eq-bc4}, one uses Slotboom variables
$\psi,u,v$.

The most important technique is to use the {\it Gummel Map} $G:H\times
H \to H\times H$, a decoupling procedure,
subsequently analyzed as a fixed point problem.
Formally, given $(u^{(k)}, v^{(k)}) \in H \times H$, one solves 
the potential equation \eqref{eq-psi} rewritten with \eqref{eq-slotboom}
\begin{multline}
\label{psi-gum}
-\nabla \cdot (\epsilon \psi^{(k+1)}) 
=
\delta^2 (\exp(-\psi^{(k+1)}) v^{(k)} \\
- \exp(\psi^{(k+1)}) u^{(k)}) + N_T,
\end{multline}
for $\psi^{(k+1)} \in H$. Then we solve the n-continuity equation
\begin{multline}
\label{n-gum}
-\nabla \cdot (D_n \exp(\psi^{(k+1)}) \nabla u^{(k+1)}) 
\\
= \frac{1}{\delta^2} R_T(\psi^{(k+1)}, u^{(k)}, v^{(k)})
\end{multline}
for $u^{(k+1)} \in H$, and the p-continuity equation
\begin{multline}
\label{p-gum}
-\nabla \cdot (D_p \exp(-\psi^{(k+1)}) 
\nabla v^{(k+1)})
\\
= \frac{1}{\delta^2} R_T(\psi^{(k+1)}, u^{(k)}, v^{(k)})
\end{multline}
for $v^{(k+1)} \in H$.  The equations \eqref{psi-gum}--\eqref{p-gum}
are supplemented with appropriate boundary conditions. 

The system \eqref{psi-gum}--\eqref{p-gum} is the iteration-lagged
system \eqref{eq-psi}--\eqref{eq-pcont} under a change of variable
formula \eqref{eq-slotboom}. The existence of a solution $(\psi,u,v)$
to the system \eqref{eq-psi}--\eqref{eq-pcont} with
\eqref{eq-slotboom} follows from first i) establishing existence of
solutions for each of the semilinear elliptic equation \eqref{psi-gum}
and the two linear elliptic equations \eqref{n-gum} and
\eqref{p-gum}. Next ii) one establishes the existence of a fixed point of
the Gummel Map. Step i) can be accomplished with standard techniques
from elliptic theory and functional analysis, while ii) the existence
of a fixed point of the Gummel Map is established from the application
of the Schauder Fixed Point Theorem, assuming that the data is small
enough. A thorough analysis of the preceeding, as well as of
regularity results, is given in \mpcite{markowich}{Sec. 3.2,3.3}.

As concerns uniqueness, one can show that under small enough external
applied voltages, the solutions are unique and depend continuously on
the data. However, under certain physical conditions, multiple
solutions to the stationary drift-diffusion model are known to
exist. e.g., for large data. A detailed exposition on uniqueness and
continuous dependence on data can be found in \mpcite{markowich}{Sec.
  3.4}. 

We note that there is no well-posedness theory available for the
heterojunction interface problem described next.

\subsection{Interface equations}
\label{sec-interface}

The first issue in an interface model is to identify which quantities
are conserved and which variables are continuous across that
interface. Some of these considerations are directly related to
parameter dependent material constants, which vary across $I$.

At a homojunction, all material parameters such as
$\eps,N_C,N_V,D_n,D_p$ are constant, and the primary variables
$\psi,n,p$, as well as their normal fluxes $\eps \nabla \psi \cdot
\nu$, $J_n \cdot \nu, J_p \cdot \nu$, are continuous. Thus the
equations \eqref{eq-psi}--\eqref{eq-pcont} hold in the classical
sense.  However, $N_T(x)$ takes a jump $\jumpz{N_T}\neq 0$.

At a heterojunction, one has to recognize two facts. First, the
material properties are not continuous across the interface.  Second,
there are physical phenomena happening in $\Omega^I$ which cannot be
described by the drift-diffusion model. Thus, in the approximate
geometrical decomposition $\Omega^I \approx I$, each of
\eqref{eq-psi}--\eqref{eq-pcont} must be replaced by a separate model
statement on $I$.  In particular, even though across $\Omega^I$ the
potential $\psi$ as well as charge densities $n,p$ are continuous,
these variables are not continuous across $I$.

The discontinuities pose issues for the mathematical model. If primary
variables are not continuous at a point $x=0$, then their derivatives
and normal fluxes across $I=\{0\}$ cannot be rigorously defined. At the
same time, we recognize that a jump of a quantity across the interface
$I$ is an artifact of geometrical approximation $\Omega^I \approx I$.
Thus one can argue on the basis of physical modeling and observation
what should be the interface equations satisfied on $I$. These
conditions on $I$ are to be understood as internal boundary conditions
that decompose the original boundary value problem on $\Omega$ into
two independent boundary value problems on $\Omega\mleft$,
$\Omega\mright$, joined by a separate interface model posed at $I$. 

Below we make the interface model on $I$ precise.  We first recall the
classical transmission conditions for a generic elliptic equation with
piecewise constant coefficients. This part is appropriate for a
homojunction and helps to set the stage for the heterojunction model
discussed next. Across a heterojunction we consider the quantities
$\psi\iright,\psi\ileft,n\iright,n\ileft,p\iright,p\ileft$ as well as
the derived quantities $(\psi_c)\iright,(\psi_c)\ileft,c=n,p$, and the
normal fluxes $\eps \nabla \psi \ileft \cdot \nu, (\nabla J_c)\ileft
\cdot \nu, c=n,p$, etc. Next we discuss the algebraic form of the
transmission problem to guide our numerical domain decomposition
formulation for heterojunctions discussed in the sequel.

To our knowledge, its mathematical and approximation properties have
not been analyzed and many are quite subtle and unusual.

\vspace*{0.2cm}
{\bf Transmission conditions for an elliptic equation at a homojunction:} 
\label{sec-transmission}
Assume that a variable $\psi$ with a flux $E=-\eps \nabla \psi$ satisfy
a second order Dirichlet boundary value problem in $\Omega$ 
\ba
\label{eq-generic}
\nabla \cdot E= -\nabla \cdot (\eps \nabla \psi) = q, \; x \in \Omega,\;
\psi\vert_{\partial \Omega}=\psi_D,
\ea
where $\eps,q,\psi_D$ are given data. (Here we use notation of the potential
equation \eqref{eq-psi} assuming that $q(x)=p(x)-n(x)+N_T(x)$ is
given.)

In order for \eqref{eq-generic} to
have a classical solution $\psi \in C^2(\Omega)$, the data $q,\eps,\eps'$ must
be continuous on $\Omega$ and in particular at $x=0$. 

Since in many practical applications when interfaces are present this
does not hold with $\jumpz{\eps}\neq 0$, $\jumpz{q}=\jumpz{N_T}\neq
0$, one considers weak solutions to \eqref{eq-generic} in which only
$\psi,E \cdot \nu$ are assumed to be continuous across $I$.  The weak
(generalized) solutions to \eqref{eq-generic} are sought in the
Sobolev space $H^1(\Omega)$ instead of in $C^2(\Omega)$, see, e.g.,
\cite{ShowDover} for details on weak solutions of elliptic problems.
In particular, if $\eps=\eps\mleft \one\mleft(x) + \eps\mright
\one\mright(x)$ is a piecewise constant coefficient, and $q=q\mleft
\one\mleft(x) + q\mright \one\mright(x)$ is a piecewise constant
source term, as long as $\min(\eps\mleft,\eps\mright)>0$, then the problem
\eqref{eq-generic} is well-posed and has a weak solution $\psi \in
H^1(\Omega)$ with $E \in L^2(\Omega)$. The well-posedness in
appropriate Sobolev spaces is a necessary condition for a proper
formulation of finite element discretizations for \eqref{eq-generic},
while (at least) $C^2$ regularity is, in general, needed for
convergence of finite difference formulations.

When solving \eqref{eq-generic} numerically, one frequently finds it
convenient to use domain decomposition (DDM) \cite{quarteroniV}. Thereby one
writes the differential equation \eqref{eq-generic} that must be
satisfied in each $\Omega\mleft,\Omega\mright$. Additionally, we write
the interface transmission conditions that need to hold at $I$. These are
\ba
\label{eq-l}
-\nabla \cdot (\eps \nabla \psi) &=& f, \; x \in \Omega\mleft,\;
\psi\vert_{\partial \Omega \cap \partial \Omega\mleft}=\psi_D,
\\
\label{eq-r}
-\nabla \cdot (\eps \nabla \psi) &=& f, \; x \in \Omega\mright,\;
\psi\vert_{\partial \Omega \cap \partial \Omega\mright}=\psi_D,
\\
\label{eq-jumpd}
\jumpz{\psi}&=&0,
\\
\label{eq-jumpz}
\jumpz{E}&=&\jumpz{\eps\nabla \psi \cdot \nu}=0.
\ea
Further information on analysis of transmission problems with
piecewise constant coefficients can be found, e.g., in
\mpcite{ShowDover}{III.4.4}.

The transmission conditions \eqref{eq-jumpd}-\eqref{eq-jumpz} describe
the qualitative nature of $\psi$, and are sufficient to close the system.
The discretized version of \eqref{eq-l}--\eqref{eq-jumpz} is uniquely
solvable and gives the same solution as the discrete version of
\eqref{eq-generic}, as long as appropriate treatment of \eqref{eq-jumpz} is 
used. 

In view of the heterojunction interface model to be developed shortly,
we remark further on the equations satisfied at $x=0$, since in
numerical point-centered formulation some equation must be posed for a
node located at $x=0$. 
We develop these equations for simplicity in 1d, by a calculus
argument. 

At a first glance it appears that the information that
\eqref{eq-generic} holds at $x=0$ is lost. To see that
\eqref{eq-generic} actually does hold at $I$, assume that \eqref{eq-l}
and \eqref{eq-r} hold pointwise, i.e., that $E'$ is continuous on each
of $\Omega\mleft,\Omega\mright$.  Then integrate \eqref{eq-r} and
\eqref{eq-l} over $(0,r)$ and $(-r,0)$, respectively, with $r$ chosen
so that these intervals are inside $\Omega$. We obtain
\ba
\label{eq-sr}
E(r) - E(0\mright)= \int_{0}^{r} q\,dx,
\\
\label{eq-sl}
E(0\mleft) - E(-r)= \int_{-r}^{0} q\,dx.
\ea
We add the two equations, divide them by $2r$, and pass to the limit
with $r \to 0$. This gives $\{E'(0)\} = \{q(0)\}$.
\ba
\label{eq-average}
\frac{E'(0\mleft)+E'(0\mright)}{2} =
\frac{q(0\mleft)+q(0\mright)}{2}.
\ea
Note that \eqref{eq-average} is derived from \eqref{eq-l}-\eqref{eq-r}
entirely independently of \eqref{eq-jumpd} and
\eqref{eq-jumpz}. 

If $q$ is continuous and \eqref{eq-jumpd}-\eqref{eq-jumpz}  hold, we get
\ba
\label{eq-generic0}
E'(0)=q(0),
\ea
thus the fact that \eqref{eq-generic} holds at $x=0$ is recovered.
Conversely, the continuity of $q$ across $x=0$ itself, without
\eqref{eq-jumpd}, \eqref{eq-jumpz}, does not guarantee that
\eqref{eq-generic0} makes sense, as the example of $\eps\equiv 1,
\psi(x)=\one\mleft(x) x^2 + \one\mright(x)(x^2+x+\pi)$ demonstrates.

We elaborate on the algebraic form of \eqref{eq-l}--\eqref{eq-jumpz}
used in numerical domain decomposition in Section~\ref{sec-dd}. 

We proceed next to define proper interface
equations for the potential and the continuity equations. 

\vspace*{0.2cm}
{\bf Potential equation at heterojunction:} Assume that
\eqref{eq-psi} holds in $\Omega\mleft$ and $\Omega\mright$, with
appropriate boundary conditions at $\partial \Omega$. To close the
system, we need to make precise the conditions at $I$, i.e., on
$\jumpi{\psi}$ and that of $\jumpi{E \cdot \nu}$. Recall that the data
$\eps,N_T$ in this equation are discontinuous at a heterojunction, and
that $q$ may not be continuous even at a homojunction.

It is known that $\psi$ is continuous across $\Omega^I$. However,
since $\psi$ must follow the energy bands, its restriction to
$\Omega_0\mleft$ and $\Omega_0\mright$ is discontinuous, and we have
\ba
\label{eq-jumppsi}
\jumpi{\psi} = \psi\iright-\psi\ileft \eqdef \mydel \psi, 
\ea
where $\mydel \psi \neq 0$ is given, see Table~\ref{tab-delta}.

The case $\jumpi{\psi}=0$ corresponds to one of the three
possibilities. First, i) either $\sigma=0$ that is, we have no
interface region $\Omega^I$ and $I$ represents a homojunction. Or, ii)
the potential is constant across $\Omega^I$, which would mean however
that the current $\eps \nabla \psi \cdot \nu$ vanishes, i.e., that $I$
is an insulating interface. The third possibility iii) is that the
potential varies in such a way across $\Omega^I$ that
$\jumpi{\psi}=0$, and this does not happen at a heterojunction.

As concerns the field $E$, it has been customary to assume that
\ba 
\label{eq-jumpe}
\jumpi{E}=\jumpi{\eps \nabla \psi \cdot\nu}=0.
\ea
It is important to comment that \eqref{eq-jumpe} describes the
shape (slope) of $\eps \nabla \psi$ at $x=-\sigma,x=\sigma$ and is not
a statement on derivatives of a function $\psi$. Since by
\eqref{eq-jumppsi} $\psi$ is discontinuous across $I$, it is not
differentiable there. However, since the width $2 \sigma$ of
$\Omega^I$ is very small, it is believed that there is no additional
net ``Dirac-delta'' charge to make $\jumpi{E}\neq 0$.

Now \eqref{eq-jumppsi} and \eqref{eq-jumpe} are sufficient to close
the system, which has a structure similar to that of DDM-like
formulations, except with a nonhomogeneous jump. We illustrate this
further in Section~\ref{sec-dd}.

Finally, it is not necessary to specify whether \eqref{eq-psi} holds {\it
  at} $I$, since any such statement should be a consequence of the
interface equations \eqref{eq-jumppsi}-\eqref{eq-jumpe} in a manner
similar to how \eqref{eq-average} was derived. However, we derive its
counterpart for modeling interest. Consider integrating
\eqref{eq-psi}, i.e., $E'=Q$, over $(-r,-\sigma)$ and $(\sigma,r)$,
for some $r >\sigma$. Assuming $E'$ is continuous on each of these
intervals, we obtain, similarly as in \eqref{eq-sr}, \eqref{eq-sl},
\bas
E\mright(r)-E\mright(\sigma)= \int_{\sigma}^{r}Q\mright dx,
\\
E\mleft(-\sigma)-E\mleft(-r) = \int_{-r}^{-\sigma}Q\mleft dx. 
\eas
Adding these equations, dividing by $r-\sigma$ 
and letting $r \to \sigma$ we obtain, 
\ba
\label{eq-psiaverage}
\frac{(E')\ileft+(E')\iright}{2} =
\frac{Q\ileft+Q\iright}{2}.
\ea
This equation, when $\sigma \to 0$, has a structure of
\eqref{eq-average} applied to \eqref{eq-psi}. 
Appropriately, if $\mydel \psi=0$, and $Q$ is continuous, we obtain
that \eqref{eq-psi} is satisfied pointwise at $x=0$.

We stress that \eqref{eq-psiaverage} is a consequence of
\eqref{eq-psi} being satisfied {\em away} from $I$, and is independent
of \eqref{eq-jumppsi}-\eqref{eq-jumpe}.  Whether or not
\eqref{eq-psiaverage} is used {\em at} $I$, depends on whether the
discrete equations are posed at $I$. Equation \eqref{eq-psiaverage} is
not significant if only weak solutions are sought.

\vspace*{0.2cm}
{\bf Continuity equation at heterojunction:}
The data $D_n$ in \eqref{eq-jn}, \eqref{eq-jp} are discontinuous
across $I$, and $R_T$ is continuous only if $n,p$ are.  Additionally,
we have \eqref{eq-jumpe}. We can still write
similarly to \eqref{eq-psiaverage} that
\ba
\label{eq-naverage}
\frac{(J_n')\iright + (J_n')\ileft}{2} = \frac{(R_T)\iright + (R_T)\ileft}{2},
\ea
which follows simply from \eqref{eq-jn} and is similar to
\eqref{eq-psiaverage}. However, we need to specify whether $n$, $J_n
\cdot \nu$ are continuous across $I$, and if not, we need a model
binding $n\ileft,(J_n)\ileft,n\iright,(J_n)\iright$. (Similar
questions concern $p,J_p$.) 

To do so, we use a model used in physical engineering literature
\cite{HorioYanai,Yang}, which does not have the same ``DDM-like''
formulation as that given for potential equation by \eqref{eq-jumppsi}
and \eqref{eq-jumpe}.  In contrast, we have an explicit interface
model for $\jumpi{J_n \cdot \nu}$ which can be interpreted as an
internal boundary condition, and is based on the notion of the {\it
  thermionic current} $J_n^I$ proportional to the jump of
$\jumpi{e^{\psi_n - E_C}}$, with a proportionality constant dependent
on the electron masses in each domain.

Early models of heterojunctions assumed $\psi_n$ is continuous across
$I$ and did not use $J_n^I$.  However, just like $\psi$, the variables
$\psi_n$ and $n$ are continuous across $\Omega^I$, but neither is
continuous across the ``idealized'' interface $I$.  From
\eqref{eq-npsin} we see that discontinuity of $\psi_n$ across $I$
parallels that of $n$, since $\jumpi{\chi}\neq 0, \jumpi{N_C}\neq
0$. Thus we have variables $n\ileft,n\iright$ that need to be related
to $J_n^I$, and this is done as in \cite{HorioYanai,Yang}
\ba
\label{eq-jnc}
J_n\cdot \nu \vert\ileft = 
J_n\cdot \nu \vert\iright = J_n^I
\eqdef a_n\mright n\iright-a_n\mleft n\ileft.
\ea
The coefficients $a_n\mright,a_n\mleft$ in \eqref{eq-jnc} are, up to
the scaling, mean electron thermal velocities, and they are calculated
depending on the temperature $T$ and on the sign of $\mydel E_C$. 

For example, let $\mydel E_C>0$ so that the conduction band jumps up a
positive amount. Then
\ba
\label{eq-idiff}
a_n\mright \eqdef \frac{A_nT^2 }{N_C\mright},\; 
a_n\mleft \eqdef \frac{A_nT^2 e^{-\mydel E_c} }{N_C \mleft}. 
\ea
Here $A_n$ is the effective Richardson's constant (effective electron
mass), and $T$ is the temperature.

If $\mydel E_c<0$, the conduction band jumps down, and we have
\ba
\label{eq-idiff2}
a_n\mright \eqdef \frac{A_n T^2 e^{\mydel E_c}}{N_C\mright},\; 
a_n\mleft \eqdef \frac{A_n T^2 }{N_C \mleft}. 
\ea

One can combine \eqref{eq-jnc}-\eqref{eq-idiff2} using quasi-Fermi energies $\psi_n$ via
\eqref{eq-npsin} to obtain
\ba
\label{eq-idiffpsi}
J_n\cdot \nu \vert\ileft = B_n\mleft(e^{(\psi_n)\ileft}
- e^{(\psi_n)\iright}),
\ea
where 
\ba
\label{eq-jnb} 
B_n\mleft \eqdef b_n\mleft e^{\psi\ileft}; \; 
b_n \mleft \eqdef A_n T^2 e^{\min (\chi\mleft,\mydel \psi + \chi\mright)}.
\ea

In the numerical calculations we use the equation \eqref{eq-idiffpsi}
instead of \eqref{eq-jnc}.

The $p$-equations are similar except with the valence band gap $\mydel
E_v$ in place of $\mydel E_c$. %
Similar to \eqref{eq-jnc}, we have
\ba
\label{eq_jpc}
J_p\cdot \nu \vert\ileft =
J_p\cdot \nu \vert\iright = J_p^I
\eqdef - a_p\mright p\iright+a_p\mleft p\ileft.
\ea
If $\mydel E_V > 0$, then the valence band jumps up a positive amount
and we have
\ba
\label{eq-pidiff}
a_p\mright \eqdef \frac{A_p T^2 e^{-\mydel E_V}}{N_V\mright},\;
a_p\mleft \eqdef \frac{A_p T^2}{N_V\mleft}.
\ea
If $\mydel E_V < 0$, the valence band jumps down and we have
\ba
\label{eq-pidiff2}
a_p\mright \eqdef \frac{A_pT^2}{N_V\mright},\;
a_p\mleft \eqdef \frac{A_pT^2 e^{\mydel E_V}}{N_V\mleft}.
\ea

It is important to note that the model defined above is appropriate at
a heterojunction only. At a homojunction with $\sigma=0$ we have
$\mydel E_c=0$, $a_n\mright =a_n \mleft$, and we expect $n \ileft =
n\iright$. Since then $J_n^I=0$, one could then infer (incorrectly)
from \eqref{eq-jnc} that $J_n \cdot \nu\ileft=0$; however, the
current $J_n$ need not vanish across $I$. Rather, at a homojunction we
have $n\ileft= n\iright$.

In summary, the interface conditions for \eqref{eq-jn} are 
\begin{multline}
\label{eq-nc}
B_n\mleft(e^{(\psi_n)\ileft}
- e^{(\psi_n)\iright})\\=
a_n\mright n\iright-a_n\mleft n\ileft= 
\left\{ \begin{array}{ll} J_n\cdot \nu \vert \ileft,&\; a_n\mright \neq a_n\mleft,
\\
0,&\; a_n\mright = a_n\mleft
\end{array}\right. ,
\end{multline}
\ba
\label{eq-jumpjn}
\jumpi{J_n \cdot \nu} &=&0.
\ea
A similar condition is formulated for the transport of holes.

We note that \eqref{eq-jumpjn} is similar to \eqref{eq-jumpe}, but
\eqref{eq-nc} is unusual, and is an internal Robin-type condition. It
is similar to external boundary conditions
\eqref{eq-bc1}--\eqref{eq-bc4} that are used for continuity equation.
Such external boundary conditions for Schottky contacts have been
defined, e.g., in \mpcite{markowich86}{Sec. 5.4}.

Mathematically, \eqref{eq-nc}-\eqref{eq-jumpjn} resemble closely the conditions that
arise for modeling fluid flow in fractures
\cite{MoralesS10,MoralesS12,Roberts05}. These are best analyzed using
a different functional setting than that in $H^1(\Omega)$. We comment
on these further in Section~\ref{sec-numerical}.

The algebraic structure of this interface problem is discussed in
Section~\ref{sec-dd}.

Finally, we mention that in some works \cite{Yang} the
continuity equation on the interface extending \eqref{eq-naverage} may
account for additional interface phenomena, i.e., {\em interface
  traps}, via additional right-hand-side interface terms active only
in $\Omega^I$, i.e., at $I$. We do not consider these here. 

\subsection{Numerical approximation}
\label{sec-numerical}

Here we discuss the discrete formulation for the subdomain models,
followed by discretization of the heterojunction interface model.

\begin{figure}[h]
\includegraphics[width=8cm]{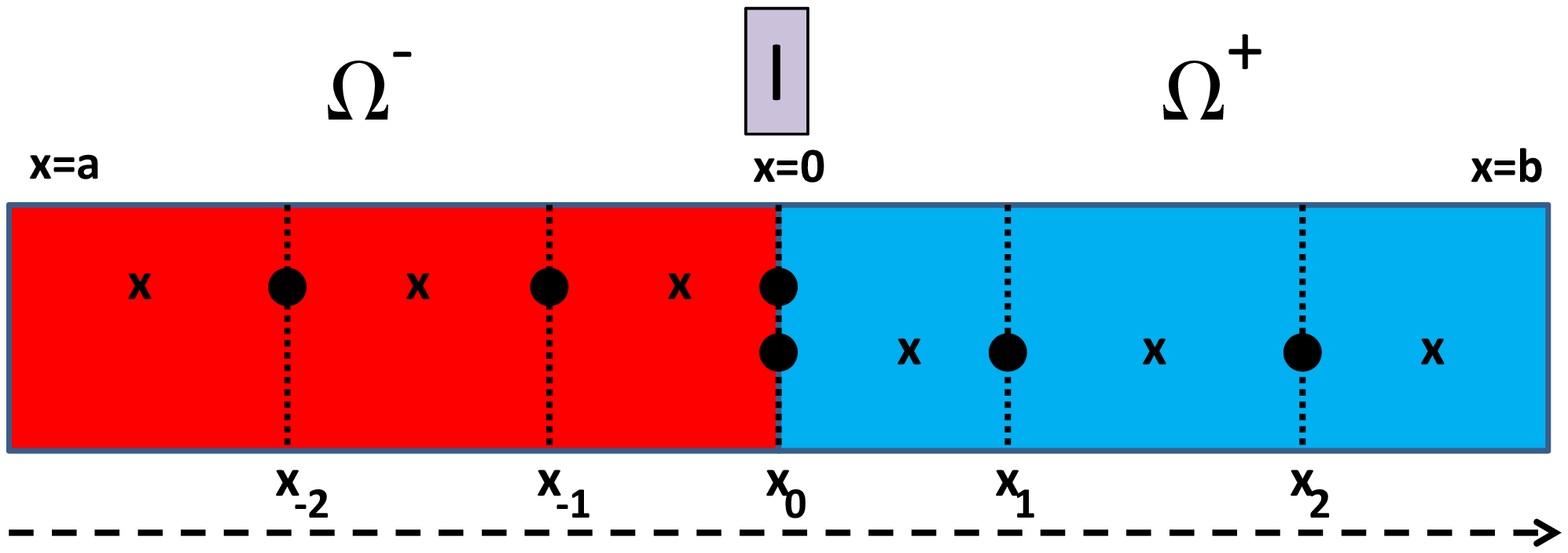}
\\
\includegraphics[width=8cm]{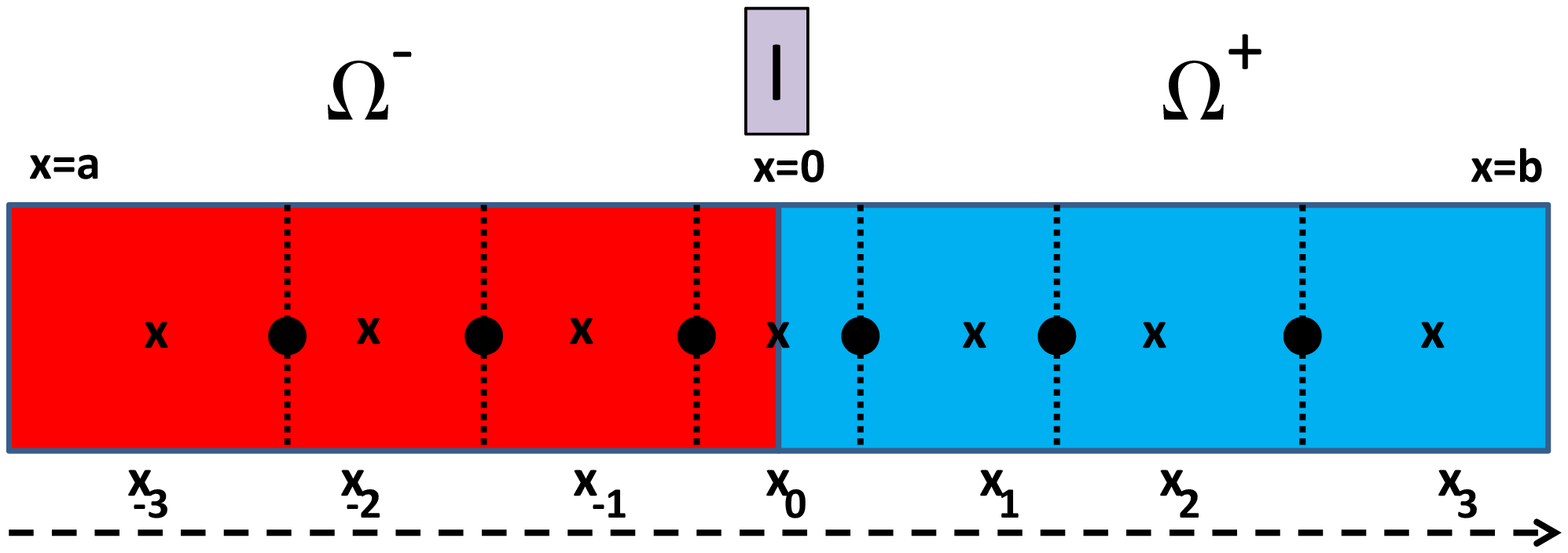}
\caption{\label{fig-grid}Grid for heterojunction domain.  {\bf Top:}
  point-centered grid used in this paper which requires doubling of
  the unknowns at the interface.  {\bf Bottom:} cell-centered grid may
  be advantageous if there is no jump in fluxes since it does not
  require doubling of the primary unknowns.
} 
\end{figure}

{\bf Grid}. We first discuss the underlying discretization.  The
equations \eqref{eq-psi}--\eqref{eq-pcont} can be discretized using
finite difference (FD) or finite element (FE) formulations, see
\cite{markowich}. In the 1d case, the FD and FE formulations are based
on a point-centered gridding of the domain, see
Figure~\ref{fig-grid}. Typically, one seeks nodal values
$\Psi_j,N_j,P_j$, approximating the primary unknowns such as
$\psi,n,p$, at grid points $x_j$ distributed in $\Omega$. For
simplicity, we consider a uniform grid with parameter $h$, and
$x_j=jh,j=0,\pm 1, \pm 2, \ldots$, as in Figure~\ref{fig-grid}. The
flux values of $E,J_n,J_p$ are approximated at $x_{j+\frac{1}{2}}$. 

Most of FD and (Galerkin) FE methods are based on point-centered
discretizations and such are those in \cite{markowich}, even though
some semiconductor modeling work
\cite{chen-cockburn1d,chen-cockburn2d} has been carried out with mixed
FE methods. We refer to \cite{BrezziFortin} for fundamental reference
to mixed FE methods, and recall that mixed FE on rectangular grids and
with lowest order spaces, and appropriate quadrature, give
cell-centered approximations \cite{RW} such as that shown in
Figure~\ref{fig-grid}, bottom. Mixed spaces were used, e.g., in the
modeling and analysis efforts in \cite{Roberts05,MoralesS12}. They
were also used extensively in numerical and domain decomposition
approaches for fluid flow where continuity of fluxes is essential,
see, e.g., \cite{GW,ACWY,PIMA00,YIMA00}. We intend to consider mixed
FE and cell-centered grids in future work, since these remove
the need for doubling interface unknowns, and may make interface
equations more natural.

In what follows we use point-centered grids.

\vspace*{0.8cm} {\bf FD formulation of the system
  \eqref{eq-psi}-\eqref{eq-pcont}.} We recall uniform point-centered
grid discretization of \eqref{eq-psi} which at node $j$, for $\eps$
constant on $(x_{j-1},x_{j+1})$ reads
\ba
\label{eq-fd}
2 \eps \Psi_j -\eps \Psi_{j-1}-\eps \Psi_{j+1}=Q_j,
\ea
with $Q_j=h^2 (P_j-N_j+N_T(x_j))$. In particular, if we assume Dirichlet
boundary conditions for the potential, \eqref{eq-fd} has the form
\ba
\label{eq-mat}
A \Psi = Q,
\ea
where $\Psi = (\ldots \psi_{-2},\psi_{-1},\psi_0,\psi_1,\psi_2
\ldots)^T$, $\frac{1}{h^2}A = -\eps \nabla^2_h$ is the tridiagonal
matrix based on the discrete 1d Laplacian $\nabla^2_h$
\cite{leveque-fdbook}. $A$ has numbers $2 \eps$ on its diagonal, and
$-\eps$ below as well as above its diagonal.  Also, $Q= (\ldots
Q_{-2},Q_{-1},Q_O,Q_1,Q_2 \ldots)^T$ is the vector of charges,
depending pointwise on $\Psi_n,\Psi_p$. Boundary conditions
\eqref{eq-bc0} are included in the right side of \eqref{eq-mat} in a standard
away \cite{leveque-fdbook}.

The challenges in the discretization of
\eqref{eq-ncont}-\eqref{eq-pcont} include the proper handling of
boundary and internal layers and of steep gradients. One uses then a
special choice of primary unknowns such as the approximations
$\Psi_n,\Psi_p$ to $\psi_n,\psi_p$, instead of approximations $N,P$ to
$n,p$. The nonlinear expressions such as those from \eqref{eq-dndp}
with \eqref{eq-npsin} must use appropriate weighting.

Consider, e.g., the electron equation \eqref{eq-ncont}, whose
discretization, with \eqref{eq-dndp}, reads
\begin{multline}
\label{eq-fdn}
(D^n_{j,j+1}+D^n_{j-1,j})\Psi_{n,j} -
D^n_{j-1,j} \Psi_{n,j-1}-D^n_{j,j+1} \Psi_{n,j+1}\\
\\
=R_j \eqdef h^2R_T(N_j,P_j),
\end{multline}
where the coefficient $D^n_{j,j+1}$ depends on
$(\Psi_{n,j},\Psi_{n,j+1})$, and typically is defined with the use of
exponential weighting, i.e., Bernoulli's function, as elegantly
described in \mpcite{BankRoseF}{eq.55},
\mpcite{markowich86}{Sec. 5.1}, \cite{Selber}. The exponential
weighting is known to help with internal boundary layers as well as to
stabilize the nonlinear solver.

Now \eqref{eq-fdn} can be written in a way similar to \eqref{eq-mat}
\ba
\label{eq-matnonl}
A(\Psi_n)\Psi_n = R(\Psi_n),
\ea
and it comprises external boundary conditions \eqref{eq-bc1},
\eqref{eq-bc3}. Here $R=(R_j)_{j=0,\pm 1,\pm 2,\ldots}$, and the
dependence of $A$ and $R$ on $\Psi,\Psi_p$ was supressed since it is
iteration lagged, as explained below. A similar equation is defined
for the transport of holes. 

In summary, in what follows we refer to the unknowns as
$\Upsilon=(\Psi,\Psi_n,\Psi_p)$, and to the equations to be solved as
\ba
\label{eq-f}
F(\Upsilon)=0,
\ea
written componentwise as 
\ba
\label{eq-npsi}
F_{\psi} (\Psi,\Psi_n,\Psi_p) =0,
\\
\label{eq-nn}
F_{n} (\Psi,\Psi_n,\Psi_p) =0,
\\
\label{eq-np}
F_{p} (\Psi,\Psi_n,\Psi_p) =0.
\ea
These correspond to the discretizations of \eqref{eq-psi},
\eqref{eq-ncont}, \eqref{eq-pcont}, respectively, with
\eqref{eq-npsin}--\eqref{eq-ppsip}, and appropriate boundary
conditions. In particular, \eqref{eq-npsi} is the same as
\eqref{eq-mat} written in residual form $F_{\psi}=A
\Psi-Q(\Psi_n,\Psi_p)$. Similarly, \eqref{eq-nn} is \eqref{eq-matnonl}
in residual form.

The system \eqref{eq-f} is solved typically with a variant of the
Newton method.
The very well known Gummel iteration is similar to that recalled in
Section~\ref{sec-analysis} solving first \eqref{eq-npsi} for
$\Psi^{(k+1)}$, given previous iteration guess, or initial guess, for
$\Psi_n^{(k)},\Psi_p^{(k)}$. Next one solves \eqref{eq-nn} for
$\Psi_n^{(k+1)}$, and \eqref{eq-np} for $\Psi_p^{(k+1)}$ using the
newly available guesses for the $\Psi^{(k+1)}$, and $\Psi^{(k+1)},
\Psi_n^{(k+1)}$, respectively. The Gummel iteration  continues
until tolerance criteria are satisfied. With a well chosen set of
unknowns, each of the component equations being solved is self-adjoint
in its primary variable. With additional iteration lagging and
appropriate choice of unknowns, each equation is linear or semi-linear
in its primary unknown.

However, as is well known, Newton's method is not globally
convergent and may fail if good initial guesses are not available.  In
practice, the cases with large gradients and ``difficult data'' are
handled with extra caution, applying, e.g., method of continuity
(homotopy) whereby one starts with small data and gradually increases
it to the desired value. See \cite{Selber} for practical information. 

Gummel iteration itself may fail at times. One situation in which this
may happen is when a spike in $E_V$ or $E_C$ forms, and intermediate
solutions cause the effective potential quantities to intersect the
spike, creating very large carrier concentrations. Recovery from this
situation can depend on the choice of variables used in the Newton
method solvers. We have seen Slotboom variables outperform
$\{\psi_n,\psi_p\}$ variables in this situation. Using the $n$ and $p$
variables has caused problems even in a single bulk material when
doping is significant. Various situation-dependent tricks have been used over
the years in the community solvers \cite{ampsweb} to control the behavior of
nonlinear iterations but it seems that a general remedy
has not yet been found.

\vspace*{0.2cm} {\bf Heterojunction}. To simulate the behavior of
$\psi,n,p$ for a heterojunction problem, one has to implement, in
addition to the discrete equations in \eqref{eq-npsi}--\eqref{eq-np}
to be solved in each of $\Omega\mleft,\Omega\mright$, the discretized
interface equations described in Section~\ref{sec-interface}. 

In particular, since $\Upsilon=(\Psi,\Psi_n,\Psi_p)$ are discontinuous
at $I$, this requires doubling the unknowns $\Upsilon$ at the grid
point $x_0=0$.  From now on this is denoted by considering
$\Upsilon\zleft, \Upsilon\zright$, with
$\Upsilon\zleft=(\Psi_0\mleft,\Psi_{n,0}\mleft,\Psi_{p,0}\mleft)$ and
$\Upsilon\zright=(\Psi\zright,\Psi_{n,0}\mright,\Psi_{p,0}\mright)$. Also,
we collect all the unknowns corresponding to $\Omega\mleft$ in
$\Upsilon\mleft$ and those for $\Omega\mright$ in $\Upsilon\mright$.
The notation is inherited by each component of $\Upsilon$.

Now each component of \eqref{eq-f} can be written as two subdomain
problems coupled to an interface problem. For example, consider
\eqref{eq-npsi} with the two subdomain problems written as
\ba
\label{eq-znpsi}
F_{\psi}\mleft (\Psi\mleft,\Psi\zleft) =0,\;\;
F_{\psi}\mright (\Psi\mright,\Psi\zright)=0,
\ea
where the dependence of $F_{\psi}\mleft$ on $\Psi\zleft$ reflects that
$\Psi\zleft$ provides the boundary values for $\Psi\mleft$, etc. Now
\eqref{eq-znpsi} must be complemented by discretization of the
interface equations so that $\Psi\mleft$ gets connected to
$\Psi\mright$. In fact, only the interface unknowns
$\Psi\zleft,\Psi\zright$, and their nearest neighbors
$\Psi_{-1},\Psi_1$ are involved in the interface model due to the
locality of the three-point FD stencil in \eqref{eq-fd}. We denote the
discrete interface problem by
\ba
\label{eq-inpsi}
F_{\psi}^I(\Psi_{-1},\Psi\zleft,\Psi\zright,\Psi_{1})=0,
\ea
and give its details in Section~\ref{sec-dd}.

A similar decomposition follows for the discrete continuity equations
\eqref{eq-nn}-\eqref{eq-np}, at each step of the Gummel
iteration.

Once the discrete equations are known, one must design the
implementation of the doubling of the unknowns as well as the
connection of the subdomain \eqref{eq-znpsi} and the interface equations
\eqref{eq-inpsi} in the solver.

In various community or commercial codes this is apparently achieved
in 1d by a monolithic approach, i.e., the interface equations are
hard-coded internally. In particular, in Gummel iteration one solves
\eqref{eq-znpsi} and \eqref{eq-inpsi} for
$\Psi=(\Psi\mleft,\Psi\zleft,\Psi\zright,\Psi\mright)$ simultaneously.
This is followed by solving for all the subdomain and interface
components of \eqref{eq-nn} for $\Psi_n$, and the loop completes with
solving \eqref{eq-np} for $\Psi_p$.

While it is perhaps easy to see how to implement these simultaneous
solves in 1d, it may be challenging or impossible to use this approach
for complicated heterojunction geometries in 2d. 

Therefore, in this paper we isolate the interface equations from the
subdomain equations with a two-fold goal.
First, we discuss an alternative approach to simultaneous solution,
based on domain decomposition method. We identify the structure of the
equations as well as pinpoint the difficulties, and this provides the
basis for future analyses and extensions.
Second, isolating the interface from the subdomains can help to handle
more general geometries, and/or implement higher order methods,
adaptive gridding and more. In particular, while for the current 1d
formulation it is easy to place the computational nodes on the
interface $I$, a general approach from the class of immersed interface
methods (IIM) \cite{GongLiLi} can be helpful to develop a solution
technique for a problem in 2d with complicated geometry. While IIM
were originally developed for problems with homogeneous jumps, there
is recent work devoted to complicated interface problems with
nonhomogenuos jumps \cite{KwakCMAME,HeLinLin}.

\subsection{Domain decomposition}
\label{sec-dd}

Domain decomposition \cite{quarteroniV} was originally designed to
accelerate or simply enable the solution of problems which were too large
to fit in a single computational core. It has since been shown to be
very effective for problems with interfaces which separate either
different materials or different physical models such as in
fluid-structure interactions or Darcy-Stokes fluid flow problems
\cite{quarteroniV}. Its mortar extensions
\cite{BerMadPat,ACWY,PIMA00,PWY02,YIMA00,LPW02} can be used to glue
together different numerical discretizations on grids that need not
match on the interface.

Domain decomposition methods (DDM) generally are iterative methods
that find the values of interface unknowns so that a proper match
between the subdomains is achieved. We refer to
\mpcite{quarteroniV}{Chap 1-2} for general background which covers DDM
for the transmission problem laid out in
Section~\ref{sec-transmission}. 

DDMs are tied to optimal solvers and preconditioners. In this context,
see \cite{Lin09} and its recent extensions which focus on
multilevel-preconditioning of non-stationary drift-diffusion systems,
without heterojunction.  The gist of the work in \cite{Lin09} is to
consider a fully coupled system solved with the Newton-Krylov
framework with multilevel preconditioning, and DDM applied here is a
purely computational technique unrelated to the presence of physical
interfaces. It will be interesting to consider in the future how to
design optimal preconditioners following \cite{Lin09} for the
heterojunction problem considered in this paper, and we hope it can
complement the domain decomposition approach of this paper which follows
naturally the material discontinuities.

In this paper we consider a non-overlapping DDM for solving an
interface problem \eqref{eq-inpsi} coupled with subdomain solvers
\eqref{eq-znpsi}.  The algorithm we outline has a double set of
interface unknowns $\Upsilon\zleft,\Upsilon\zright$ instead of a
single $\Upsilon_0$ as in the traditional set-up.  It handles
nonhomogeneous jumps of the primary unknowns and can also handle
nonhomogeneous jumps of the flux(es).  We have applied it successfully
to the solution of the potential equation at both homo- and
heterojunctions, and to the continuity equation at a homojunction. The
interface iteration for continuity equations at a heterojunction is
currently in progress.

\subsubsection{DDM for the potential equation}

We discuss first DDM for the algebraic problem that corresponds to the
discretized form of \eqref{eq-mat} for the potential equation at
a homojunction, i.e., with a single interface unknown
$\Psi_0$. Ordering the unknowns so that
$\Psi=(\Psi\mleft,\Psi\mright,\Psi_0)^T$ we have the following system
\ba
\label{eq-dd}
\left[ 
\begin{array}{ccc}
A_{\ml,\ml} & &A_{\ml,0}\\
& A_{\mr,\mr} &A_{\mr,0}\\
A_{0,\ml} &A_{0,\mr}&A_{0,0}\\
\end{array}
\right]
\left[ 
\begin{array}{c}
\Psi\mleft\\
\Psi\mright\\
\Psi_0
\end{array}
\right]
=
\left[ 
\begin{array}{c}
Q\mleft\\
Q\mright\\
Q_0
\end{array}
\right],
\ea
which has the classical form (up to notation) from
\mpcite{quarteroniV}{Sec 2.3}.  Here $A_{\ml,\ml}$ is the part of $A$
matrix corresponding to the interior nodes of $\Omega\mleft$, and
$A_{\ml,0}=A_{0,\ml}^T=[\ldots -\eps]^T$ represents the coupling between
the nodes in $\Omega\mleft$ and those at $I$. Also,
$A_{\mr,0}=A_{0,\mr}^T=[-\eps \ldots]^T$, while $A_{0,0}=2 \eps$ is just a
number, i.e., a $1 \times 1$ matrix. etc. 

The DDM is an iterative method for solving the system in the
Schur-complement form
\ba
\label{eq-schur}
\Xi \Psi_0 = \Theta,
\ea
where we obtain by block elimination, e.g., 
\ba
\label{eq-block}
A_{\ml,\ml} \Psi\mleft=
Q\mleft - A_{\ml,0} \Psi_0,
\ea
from \eqref{eq-dd} that $\Xi := A_{0,0} - A_{0,-} A_{-,-}^{-1} A_{-,0}
- A_{0,+} A_{+,+}^{-1} A_{+,0}$, and $\Theta := Q_0 - A_{0,-}
A_{-,-}^{-1} Q^- - A_{0,+} A_{+,+}^{-1} Q^+$.  The problem
\eqref{eq-schur} has a simple structure thanks to linearity of
\eqref{eq-dd}. See Section~\ref{sec-appendix-ddh} for the calculations of
$\Xi,\Phi$ which includes \eqref{eq-schur} as a special case.

It is well known that one does not form $\Xi$ explicitly. Rather, we
use its structure and properties in an iterative solver for
\eqref{eq-schur}, which subsequently only requires subdomain
solvers. In particular, an iterative solver delivers quesses
$\Psi_0^{(k)},k=1,2,\ldots$, and requires that we compute a
matrix-vector product $\Xi \Psi_0^{(k)}$ for a given guess
$\Psi_0^{(k)}$. This in turn requires that we evaluate, e.g., $A_{0,-}
A_{-.-}^{-1} A_{-,0} \Psi_0^{(k)}$, where $A_{-.-}^{-1}$ is not needed
explicitly. Rather, a linear system with $A_{-.-}^{-1}$ is solved, and
this corresponds to solving a problem on the subdomain $\Omega\mleft$
using the boundary conditions on $I$ provided by $\Psi_0^{(k)}$.  A
new guess $\Psi_0^{(k+1)}$ is computed depending on the residual of
\eqref{eq-schur}; the details depend on the choice of interface
solver, see Section~\ref{sec-ddresults}.

\vspace*{0.2cm} {\bf Equivalence to the transmission problem and
  doubling interface unknowns for homojunction.}
One can easily show that \eqref{eq-dd} is equivalent to discretizing
\eqref{eq-generic} in the transmission form
\eqref{eq-l}-\eqref{eq-jumpz}. First,
we recognize that the first two block rows of
\eqref{eq-dd} are the discrete counterparts of
\eqref{eq-l}-\eqref{eq-r}. Each includes the coupling to the
interface unknown $\Psi_0$ used as a boundary condition. 
We can also formally double the unknowns and replace
$\Psi_0$ by $\Psi_0\mleft,\Psi_0\mright$ on the interface. Then we
enforce \eqref{eq-jumpd} explicitly by setting them equal to each
other.

With doubling of the unknowns on the interface the system \eqref{eq-dd} becomes
\ba
\label{eq-dd2}
\left[ 
\begin{array}{cccc}
A_{\ml,\ml} & &A_{\ml,0}&\\
& A_{\mr,\mr} &&A_{\mr,0}\\
A_{0,\ml} &A_{0,\mr}&\frac{1}{2}A_{0,0}&\frac{1}{2}A_{0,0}\\
&&-1&1\\
\end{array}
\right]
\left[ 
\begin{array}{c}
\Psi\mleft\\
\Psi\mright\\
\Psi\zleft\\
\Psi\zright
\end{array} 
\right]
=
\left[ 
\begin{array}{c}
Q\mleft\\
Q\mright\\
Q_0\\
0
\end{array}
\right].
\ea
The last row of \eqref{eq-dd2} expresses \eqref{eq-jumpd}, thus we can
eliminate one of the two values $\Psi\zleft,\Psi\zright$, and the
system reduces to \eqref{eq-dd}. Note that we have intentionally mixed
the symbols for matrices with those for numbers in the last row.

The second row from below, (and equivalently, the last row in
\eqref{eq-dd}), can be shown to follow from the FD discretization of
\eqref{eq-jumpz} using a second order accurate formula on each side of
$x_0=0$, followed by a discrete equation \eqref{eq-fd} to be satisfied
at $j=0$ involving ghost nodes. See Section~\ref{sec-appendix-dd} for
details.

\vspace*{0.2cm} 
{\bf Accounting for the jump of potential on the interface for heterojunction.} Now we
extend \eqref{eq-dd2} and define the algebraic problem arising in the
potential part of the Gummel iteration \eqref{eq-npsi} with
a heterojunction. We have a given $Q=Q^{(k)}$ and we need to solve an
appropriate counterpart of \eqref{eq-fd} for $\Psi=\Psi^{(k+1)}$.

With the notation as above, we place a computational node at $x_0$,
and seek approximations $(\Psi\zleft,\Psi\zright)$ to
$(\psi\ileft,\psi\iright)$ so that discretized versions of
\eqref{eq-jumppsi} and \eqref{eq-jumpe} are satisfied. In addition, we
modify the matrices in \eqref{eq-dd2} to account for the values of
$\eps\mleft,\eps\mright$. We also modify the right hand side in the
last row of \eqref{eq-dd2} to account for the nonhomogeneous jump in
$\psi$ across $I$, and we let $\avei{Q^I}$ replace $Q_0$ per
\eqref{eq-psiaverage}. 

We obtain
\ba
\label{eq-ddh}
\left[
\begin{array}{cccc}
A_{-,-} & &A_{-,0}&\\
& A_{+,+} &&A_{+,0}\\
A_{0,-} &A_{0,+}&A_{0,0}\mleft&A_{0,0}\mright\\
&&-1&1\\
\end{array}
\right]
\left[
\begin{array}{c}
\Psi^-\\
\Psi^+\\
\Psi^-_0\\
\Psi^+_0
\end{array}
\right]
=
\left[
\begin{array}{c}
Q^-\\
Q^+\\
\avei{Q^I}\\
{\mydel \psi}
\end{array}
\right].
\ea
We refer to Section~\ref{sec-appendix-dd} for details. Now 
if $\mydel \psi=0$, then \eqref{eq-ddh} reduces to \eqref{eq-dd2} and
further to \eqref{eq-dd}, if also $Q\zleft=Q\zright$.

One can write an explicit calculation to set up an interface
problem similar to \eqref{eq-schur} 
\ba
\label{eq-schur-ddh}
\Xi \Psi_0^- = \Phi.
\ea 
The matrix $\Xi$ now accounts for different values of $\Psi^-_0,\Psi^+_0$ on the
interface and $\Phi$ includes $\mydel \psi$.

We have 
\begin{multline}
\label{Xi-heterdd}
\Xi = A_{0,0}^- + A_{0,0}^+ - A_{0,-} A_{-,-}^{-1} A_{-,0} \\- A_{0,+} A_{+,+}^{-1} A_{+,0},
\end{multline}
\begin{multline}
\label{Phi-heterdd}
\Phi = \left(A_{0,-} A_{-,-}^{-1} Q^- + A_{0,+} A_{+,+} Q^+ + \avei{Q^I}\right)
\\+ (A_{0,0}^+ - A_{0,+} A_{+,+}^{-1}) \mydel \psi.
\end{multline}
See Section~\ref{sec-appendix-ddh} for details of this calculation and
Section~\ref{sec-ddresults} for the iteration to solve \eqref{eq-schur-ddh}.

\subsubsection{DDM for continuity equations}

Next we proceed to define the algebraic problem corresponding to
\eqref{eq-nn}, with particular attention paid to interface equations
at homo- and heterojunctions. A formulation for \eqref{eq-np}
can be written similarly.

{\bf Homojunction.} First we see that we can rewrite \eqref{eq-fdn} in
a DDM form for the homojunction similarly to \eqref{eq-dd2} written for
\eqref{eq-fd} with the matrix $A$ calculated using the coefficients
$D^n_{j,j+1}$ instead of the constant $\eps$. The unknowns are
$(\Psi_n\mleft,\Psi_n\mright,\Psi_{n,0}\mleft,\Psi_{n,0}\mright)^T$,
and the right hand side vector is now
$R^T=(R\mleft,R\mright,R_0,0)^T$, where the last entry follows from
the continuity of $\psi_n$ at a homojunction, reflected in the discrete
problem by $\Psi_{n,\ml}^0=\Psi_{n,\mr}^0$.

The structure of the appropriate algebraic problem to be solved for
$\Psi_n$ is thus entirely analogous to \eqref{eq-dd2}.  The major
difference with respect to \eqref{eq-dd2} is that the components $R_j$
of the vector $R$ depend nonlinearly on the unknowns $\Psi_{n,j}$. The
same concerns the coefficients
$D^n_{j,j+1}=D^n_{j,j+1}(\Upsilon_j,\Upsilon_{j+1})$, thus
$A=A(\Psi_n)$.

The nonlinearity does not change the structure of the analogue of
\eqref{eq-dd} for homojunction, but a simple reformulation with a
Schur-complement as in \eqref{eq-schur} is no more possible, since
block elimination is not available and, e.g.,
\begin{multline}
\label{eq-blockpsin}
A_{\ml,\ml}(\Psi_n,\Psi_{n,0}\mleft) \Psi_n\mleft\\=
R\mleft - A_{\ml,0} (\Psi_n,\Psi_{n,0}\mleft) \Psi_{n,0},
\end{multline}
replaces \eqref{eq-block}. However, at least theoretically, we can
derive the nonlinear counterpart of \eqref{eq-schur} from
\eqref{eq-blockpsin}
\ba
\label{eq-nschur}
\Xi_n(\Psi_n)=\Phi_n.
\ea
After linearization (or in a Newton step, or via Gummel
iteration-lagging), one can identify the domain decomposed blocks of the
nonlinear system analogous to those in \eqref{eq-schur}.

\vspace*{0.2cm}
{\bf Heterojunction.}
We now outline the DDM version of the heterojunction system in
analogy to \eqref{eq-ddh}. 

We have \eqref{eq-jumpjn} similarly to \eqref{eq-jumpe}, therefore the
third row of our system will look alike \eqref{eq-ddh}. We write 
it properly as (see also \eqref{eq-jall} in Appendix)
\begin{multline}
\label{eq-psinall}
D^n_{-1,0\mleft}\Psi_{n,0}\mleft + D^n_{0\mright,1} \Psi_{n,0}\mright
\\-D^n_{-1,0\mleft}\Psi_{n,-1}-D^n_{0\mright,1}\Psi_{n,1}=\avei{R^I},
\end{multline}
and recall that the coefficients $D^n_{-1,0\mleft}$ depend nonlinearly
on the values of $\Upsilon_{-1}$ and $\Upsilon_{0\mleft}$.  In the
block form we have therefore 
\begin{multline}
A_{0,-} \Psi_{n}\mleft +
A_{0,+} \Psi_{n}\mright +
A_{0,0}\mleft \Psi_{n,0}\mleft +
A_{0,0}\mright \Psi_{n,0}\mright\\ =\avei{R^I},
\end{multline}
where we identify $A_{0,0}\mleft=D^n_{-1,0\mleft}$,
$A_{0,0}\mright=D^n_{0\mright,1}$ and $A_{\ml,0}=A_{0,\ml}^T=[\ldots
  -D^n_{-1,0\mleft}]^T$, and other definitions can be completed
similarly to those done for \eqref{eq-dd}, \eqref{eq-dd2} and
\eqref{eq-ddh}.

Next, instead of a Dirichlet-like condition \eqref{eq-jumppsi} which
has a simple discretization in the last row of \eqref{eq-ddh}, we have
the Robin-like condition \eqref{eq-nc}. We discuss its discretization
and structure next.

In order to discretize \eqref{eq-nc} to second order accuracy, we
should use the ghost variables as in the derivation leading to
\eqref{eq-dd2}, as is done carefully in Section~\ref{sec-appendix} for
the potential equation. An alternative is to use a first order, one
sided approximation to $J_n\vert\ileft$; see
Section~\ref{sec-appendix-dd} and \eqref{eq-j0} for details on
one-sided derivatives at the interface.

However, by \eqref{eq-psin} we have to deal with nonlinear dependence
of $D^n$ on the variables involved. Thus, even though the first order
approximation is less accurate, it appears also to be the most
straightforward. We describe it here and leave the more accurate
formulation for future work.

We approximate $J_n\vert\ileft \approx D^n_{-1,0\mleft}
\frac{\Psi_{n,0}-\Psi_{n,-1}}{h}$ and set as the discrete counterpart of \eqref{eq-nc}
\bas
D^n_{-1,0\mleft} \frac{\Psi_{n,0}\mleft-\Psi_{n,-1}}{h} = B_n\mleft 
(e^{\Psi_{n,0}\mright}-e^{\Psi_{n,0}\mleft}),
\eas
and a simple reformulation gives us finally
\begin{multline}
\label{eq-jnc0}
D^n_{-1,0\mleft} (\Psi_{n,0}\mleft-\Psi_{n,-1}) \\
-hB_n\mleft 
e^{\Psi_{n,0}\mright}+hB_n\mright e^{\Psi_{n,0}\mleft}=0,
\end{multline}
where the left hand side of \eqref{eq-jnc0} depends nonlinearly on
$\Psi_{n,0}\mright,\Psi_{n,0}\mleft$, and on
$\Psi_{n,-1},\Psi_{n,0}\mleft$ via $D^n_{-1,0\mleft} =
D^n_{-1,0\mleft}(\Psi_{n,-1},\Psi_{n,0}\mleft)$.

We are then ready to state the analogue of \eqref{eq-ddh} for the
electron continuity \eqref{eq-nn} equation
\ba
\label{eq-ddn}
\left[
\begin{array}{cccc}
\!A_{-,-} & &A_{-,0}&\\
& A_{+,+} &&A_{+,0}\!\\
\!A_{0,-} &A_{0,+}&A_{0,0}\mleft&A_{0,0}\mright\!\\
\end{array}
\right]
\!\!\left[
\begin{array}{c}
\Psi_n\mleft\\
\Psi_n\mright\\
\Psi_{n,0}\mleft\\
\Psi_{n,0}\mright
\end{array}
\right]
&=&
\left[
\begin{array}{c}
\!R^-\!\\
\!R^+\!\\
\!\avei{R}\!\\
\end{array}
\right],
\\
\nonumber
A_{0,\mleft} \Psi\mleft + g\mleft(\Psi_{n,0}\mleft)
+g\mright(\Psi_{n,0}\mright)&=&0,
\ea
where the definitions $g\mleft,g\mright:\R \to \R$ follow directly 
from \eqref{eq-jnc0}.

One could formally eliminate the (block) unknowns in
\eqref{eq-ddn} and set-up the nonlinear Schur-complement formulation
extending further the nonlinear homojunction case \eqref{eq-nschur}
and the linear heterojunction case \eqref{eq-schur-ddh} to
\ba
\label{eq-nschurhet}
\Xi_n(\Psi_{n,0}\mleft)=\Phi_n.
\ea
Note that \eqref{eq-nschurhet} involves untangling of the nonlinear
relationship between $\Psi_{n,0}\mleft$ and $\Psi_{n,0}\mright$ from
the last row in \eqref{eq-ddn}, as well as other nonlinear
relationships from other rows similarly to those in
\eqref{eq-blockpsin}. We won't pursue the explicit form of
$\Xi_n$ or of $\Phi_n$ from \eqref{eq-nschurhet} since these are 
not needed by the
actual iterative solver for \eqref{eq-nschurhet}. While we have
successfully solved \eqref{eq-nschur} with an algorithm similar to that
for \eqref{eq-schur}, and have some preliminary results on solving
\eqref{eq-nschurhet}, details will be given elsewhere.

\section{Results and examples}
\label{sec-results}

In this section we present computational simulations of processes at
heterojunctions, and in particular those supporting the search for
more efficient solar cells. 

In Section~\ref{sec-dfresults} we show DFT and continuum model results
for \matsiz\ which emphasize the two scales present in the problem. In
Section~\ref{sec-cigs} we present results of a continuum model for a
common solar cell \matcigs\ heterojunction, and demonstrate the
photocurrent. In Section~\ref{sec-ddresults} we focus on the DDM
solver and study its performance, first for a Si homojunction, and
next for heterojunctions. Here we consider three heterojunctions:
silicon-galium arsenide (\matsig), silicon-zinc sulfide (\matsiz), and
copper indium galium selenide-cadmium sulfide (\matcigs). (As concerns
CIGS, we use the alloy semiconductor CuIn$_{1-x}$Ga$_{x}$Se$_2$, where
$x$ is the ratio of the number of atoms of Ga to that of Ga plus
In. Typically, $x \approx 0.3$ for thin film solar absorbers.).

Since the DDM for the continuity part of the heterojunction model
\eqref{eq-nn}-\eqref{eq-np} is still under development, we use the
monolithic solver for the continuum model in
Sections~\ref{sec-dfresults}-\ref{sec-cigs}.

\medskip

We use $\Omega\eqdef (a,b)=(-.1,1) \mu${m} for \matsiz\ in
Section~\ref{sec-dfresults}, also $\Omega=(a,b)=(-.2,2) \mu${m} in
Section~\ref{sec-cigs}, and $(a,b)=(-1,1) \mu${m} in all DDM examples
in Section~\ref{sec-ddresults}. The interface in all sections is
always at $x=0$. All calculations are assumed to be done at ``room
temperature'' or $300$ K, and material constants are listed in
Section~\ref{sec-df_appendix}.

In Sections~\ref{sec-dfresults} and ~\ref{sec-cigs} we
use the conventional band diagrams depicting the energy levels for the
electron $E_{vac},E_C$, and $E_V$ defined in \eqref{eq-vac}.
Also, at thermal equilibrium, we have
$
E_{Fn} = E_{Fp} = E_F,
$
where the Fermi level $E_F$ is defined only in thermal equilibrium, and is
always constant in a single body or device throughout which electrons
may move to reach an equilibrium distribution.  Without loss of
generality we set $E_F = 0$, which is consistent with
what was done in Section~\ref{sec-bc}.

\subsection{Results of DFT and continuum models 
for \matsiz\ interface}
\label{sec-dfresults}

\begin{figure}[h]
\includegraphics{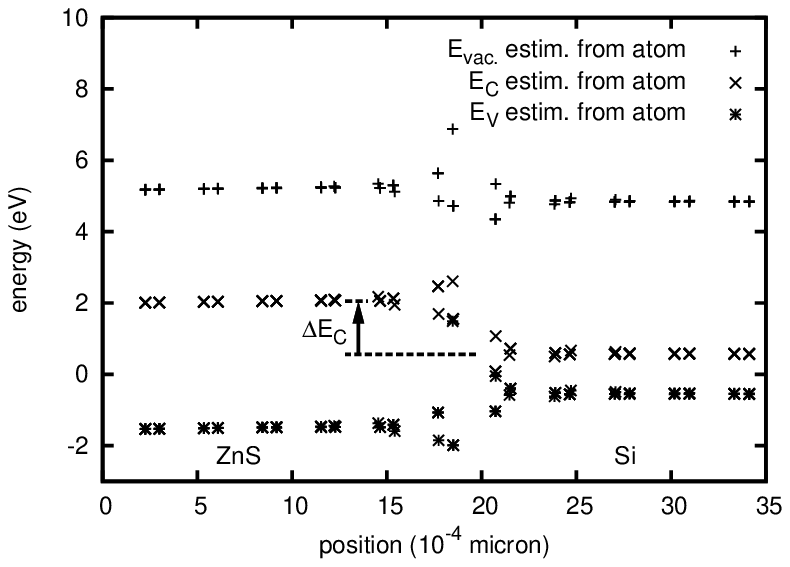}
\\
\includegraphics[width=8cm]{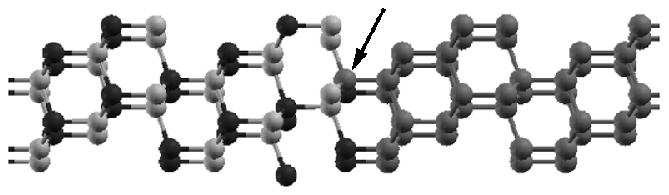}
\\
\includegraphics{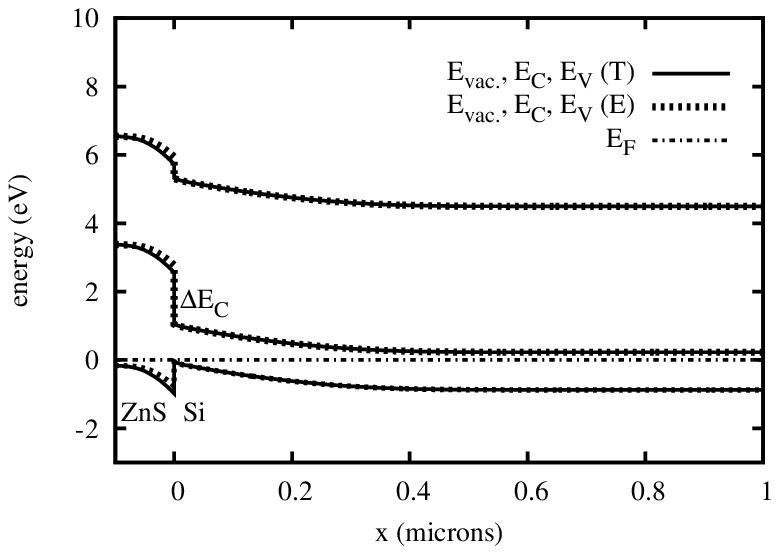}
\caption{\label{fig-SiZnS}Microscopic and macroscopic band behavior of the Si/ZnS interface (ZnS on left).
{\bf Top:} Physical electron energy bands traced by the DFT-calculated potential spatially averaged over individual atomic spheres (points show atom centers).
The three bands are obtained by adding offsets derived from $\chi$, $E_g$, and results from single material DFT calculations.
{\bf Center:} Atoms in the ion-relaxed supercell within approximately the same $x$ range. Zn atoms are the lightest; S atoms are darkest. The substituting defect is pointed to by an arrow.
{\bf Bottom:} Bands at thermal equilibrium, (T)heory and (E)xperiment.
}
\end{figure}

Here we first discuss the DFT model calculation of $\Delta E_C$ for
\matsiz. Next we compare the results of the continuum model using this
value and an available experimental value. The results are illustrated
together in Figure~\ref{fig-SiZnS}, where the post-processed DFT
results over $\Omega^I$ are shown along with atomic structure of
$\Omega^I$, followed by results of a continuum model where, as we explained
in Section~\ref{sec-intro}, $\Omega^I \approx I$.

Our study of both DFT and continuum models of the \matsiz\ interface is
motivated by fundamental interest in polar interfaces
\cite{harrison78pol,nakagawa06why} as well as by the hypothesis of
HAII, as discussed in Section~\ref{sec-intro}.  Our DFT model examines
atomically distinguishable \matsiz\ interfaces having the (111) normal
orientation
of energetically stable interface defects.  
A quantum mechanical and electrostatic analysis of single atom defect
variations for other crystal orientations (111), (100), and (011) in \matsiz\
will be presented elsewhere \cite{foster13unp}.

The DFT calculations following the model discussed in
Section~\ref{sec-dft} show that a particular stable interface defect,
the replacement of one quarter of the S atoms at an atomically abrupt
Si-S interface by Si atoms (see Figure~\ref{fig-SiZnS}), yields a
conduction band offset $\Delta E_C \approx 1.5$ eV in reasonable
agreement with the experimental result $1.73 \pm 0.2$ eV from
experiment \cite{maierhofer91val}
\ba
\Delta E_C = E_C\vert_{ZnS}\mleft - E_C\vert_{Si}\mright = 1.73 \pm 0.2 \textrm{ eV.}
\ea
The result is not fully predictive, as an alternative single atom
substitution defect at the Si-S interface is also found to be
energetically stable, yet has $\Delta E_C \approx 0.3$ to $0.4$ eV.
The calculated energy required to form these two interfaces are within
the spread of the calculated data ($\approx \pm 0.1$ eV), and thus in
this case one cannot determine from the DFT calculation alone which
interface is more likely to form.

The top two parts of Figure~\ref{fig-SiZnS} make clear the atomic bond
length scale of the interface region $\Omega_I$ over which the bands
energies change.  Each atom contributes an estimate for the local
position of each of the three electron energy levels.  Also, the
conduction band offset of approximately $1.5$ eV is denoted by the
vertical arrow.  The bottom portion of Figure~\ref{fig-SiZnS} shows
results of the continuum model using the theoretical (T) and
experimental (E) band alignment parameters. Here $T$ refers to 
the computational simulations for the DFT. The model simulates the
macroscopic band bending at thermal equilibrium, with
$N_T\mleft\vert_{ZnS} = - 10^{17}$ cm$^{-3}$, and
$N_T\mright\vert_{Si} = 5\times 10^{15}$ cm$^{-3}$ in Si. We see that
the difference between experimental (E) and theoretically (T) predicted
values is small.

As concerns the results of DFT, we note that an overall slope that is
an artifact of $x$ periodicity in the DFT calculation has been
removed.  The small difference in the slopes apparent between the two
sides of the interface also arises from artifacts due to the $x$
periodicity in the DFT calculation.

The band slopes indicate that an electron in the ZnS conduction band
will drift toward the interface.  In most regions the electrons will
drift right while the holes will drift left.  However, the valence
band offset of magnitude 0.7 eV \cite{maierhofer91val} to 0.9 eV (our
calculation) serves as a barrier to the left-bound holes.  Using a
uniform carrier generation $G$ to represent solar absorption, the
model yields a physically small photocurrent ($< 10^{-4}$ A/cm$^2$)
for this 1D device.  This will likely not be a problem for solar cell
design involving 2d or 3d nanostructures.

\subsection{Results of the continuum model for the \matcigs\ solar cell}
\label{sec-cigs}

Here we show the continuum model results for the solar cell heterojunction
\matcigs, with data for the band offsets from \cite{minemoto01the,
  gloeckler03num, wei93ban}. We show simulation results and
describe several quantities of interest that arise from such
simulations.

\begin{figure}[h]
\includegraphics{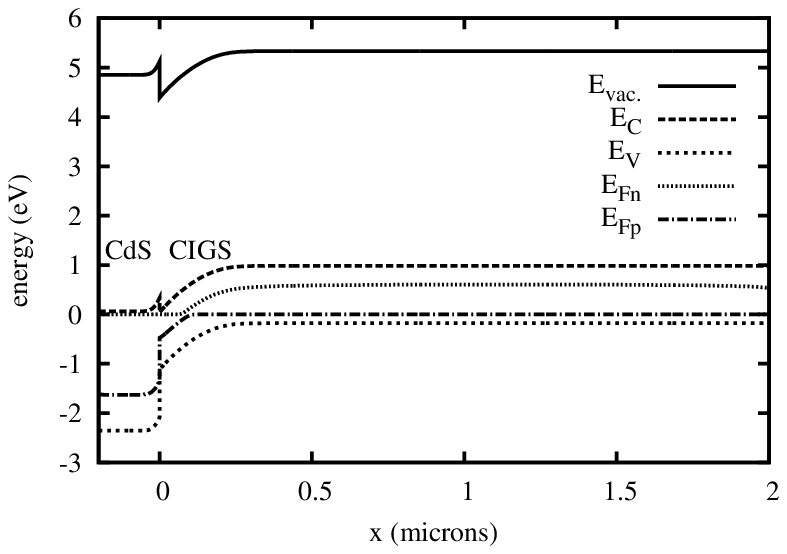}
\\
\includegraphics{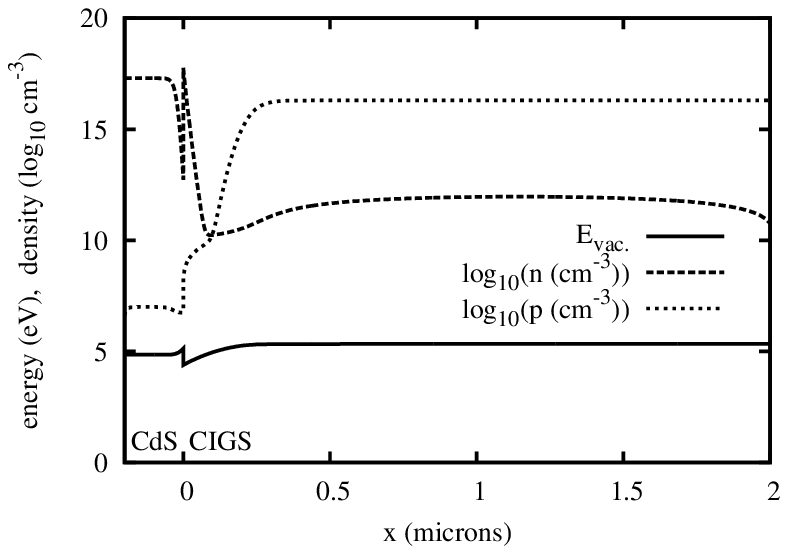}
\\
\includegraphics{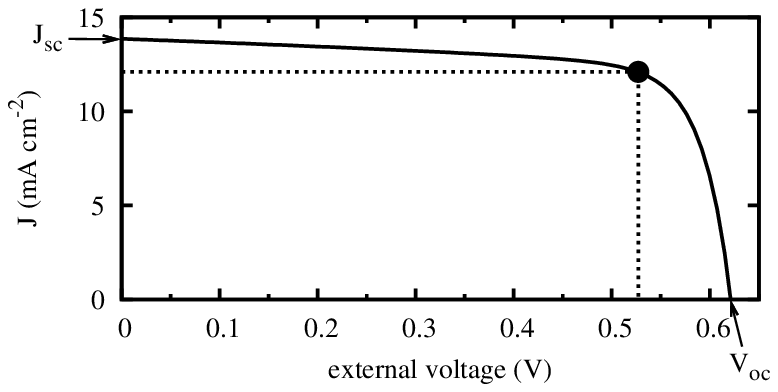}
\caption{\label{fig-CIGSCdSshortcircuit}The CIGS/CdS heterojunction under illumination (CdS on left).
{\bf Top:} Band diagram with effective Fermi potentials.
{\bf Center:} $\log_{10}$ of the carrier densities and $E_{vac}$.
{\bf Bottom:} $J$-$V$ curve showing current per solar cell area versus external voltage.
(Current is given in milliamps (mA).)
The black circle denotes the parameters yielding maximum power.
}
\end{figure}

Figure~\ref{fig-CIGSCdSshortcircuit} shows the simulation for the
``short circuit'' case in which the device is illuminated and
\ba
\label{eq-shortcircuit_cond}
V_a = V_b  = 0.
\ea
See Section~\ref{sec-bc} for context. 
The doping profile or
fixed charge profile is set to be $N_T\vert_{CdS} \mleft = 2 \times 10^{17}$ cm$^{-3}$
and $N_T\vert_{CIGS} \mright = -2\times 10^{16}$ cm$^{-3}$.

Also, we use a simple piecewise constant $G=10^{18}\one\mleft +10^{21}\one\mright$
electron-hole pairs generated per cm$^3$s. (In practice, this model
$G$ would be replaced by an optical absorption profile calculated
using 1D optics, the wavelength-dependendent absorptivity of the
materials, and the spectrum of sunlight.)  

The top and center parts of Figure~\ref{fig-CIGSCdSshortcircuit} show
simulation results. In particular, it is clear that $n,p$ exhibit very
large relative and absolute discontinuities at the interface.

Next we discuss the quantities of interest for heterojunction
simulations.  The bottom part of Figure~\ref{fig-CIGSCdSshortcircuit}
shows the $J$-$V$ curve, in physical units, of the rounded rectangular shape
characteristic of solar cell devices \cite{Sze}.  The $J$-$V$ curve
identifies the efficiency of a solar cell. In the $J$-$V$ curve the
values $J=J_n+J_p$ are computed from \eqref{eq-f}  for several
values of 
$
V=V_b - V_a
$ for which we run simulations.

Finally, the current flowing through a circuit powered by a solar cell of
area $A$ is given by 
\ba
\label{eq-IJA}
I_A = J A.  \ea 
The current at $V=0$ is known as the short circuit current,
and $I_{\textrm{sc}}$ is identified in Figure~\ref{fig-CIGSCdSshortcircuit}.
In practice, the voltage across the external circuit is varied by a
resistance or load $R_L$ placed in an external circuit, 
\ba
\label{eq-VIR}
V = I_A(V) R_L .
\ea
Increasing the resistance raises $V$, and it lowers the current that
flows in the circuit, thus as $R_L \rightarrow \infty$, $I_A
\rightarrow 0$.  The voltage at $I_A=0$ is known as the open circuit
voltage, $V_{\textrm{oc}}$, also shown in the Figure.
The values $I_{\textrm{sc}}$ and $V_{\textrm{oc}}$ are respectively
the maximum current and maximum voltage that a system (illumination +
device) can produce.  

The power delivered by the solar cell to the external portion
of the circuit is the product
\ba
P_A = I_A V,
\ea
and the maximum power $P_{\textrm{max}}$ on the $J$--$V$ curve is denoted by
the black circle.  The ratio of $P_{\textrm{max}}$ to the product
$J_{\textrm{sc}} V_{\textrm{oc}}$ is a figure of merit known as the
fill factor $F_F$.  While the fill factor depends on the details of
optical absorption, the flat $G$ modeled here yields a fill factor of
$F_F = 0.74$.  This compares well with real systems, in which $F_F$
values greater than 0.8 are considered good \cite{Sze}.

\subsection{Results of continuum model with DDM}
\label{sec-ddresults}

In this Section we present results on DDM applied to 1d
drift-diffusion system \eqref{eq-f} for homojunction and
heterojunction examples.  Recall that we use quasi-Fermi variables and
the Gummel decoupling method as explained in
Section~\ref{sec-numerical}. We first specify the algorithm used for
solving the interface problems, and next present results. We are
interested in how the algorithm compares to the monolithic solver, and
how to tune its performance. Furthermore, we check for
mesh-independence and accuracy, i.e, grid convergence.

For homojunction, we have applied DDM in a Gummel iteration to each
component equation of \eqref{eq-f}, i.e., to
\eqref{eq-npsi}--\eqref{eq-np}, where each of the variables other than
the primary is iteration-lagged. 

For heterojunction, we present results only for the potential equation
\eqref{eq-npsi}. While preliminary results on
\eqref{eq-nn}--\eqref{eq-np} are promising, we do not show them here.

We denote by $N$ the number of nodes of the grid, and $h =
\frac{b-a}{N-1}$ is the grid parameter for the monolithic solver and
$h=\frac{b-a}{N-2}$ for heterojunction examples.

We emphasize that the use of DDM for potential equation appears
straightforward but it applies to a semilinear 1d problem (with
iteration-lagged variables). Moreover, the system \eqref{eq-f}
is nonlinear. Thus, tuning its performance is delicate especially for
the heterojunction case. The use of DDM for the continuity equation is
tricky even for the 1d homojunction case since it involves nonlinear
equations.

\vspace*{.2cm} {\bf Iterative algorithm on interface.} We first present the
algorithm for the potential equation at a homojunction.  Recall
that the transmission conditions \eqref{eq-jumpd}-\eqref{eq-jumpz}
translate in algebraic form to the DDM framework in \eqref{eq-dd}. To
solve algebraically \eqref{eq-schur}, we iterate taking guesses for
$\Phi_0=\lambda^{(k)}$ as follows.

{\it Algorithm (NN) for \eqref{eq-jumpd}-\eqref{eq-jumpz}} Given some
$\lambda^{(0)}$ and a parameter $\theta>0$, proceed
iteratively for $k=0,1,\ldots$
\begin{enumerate}
\item Solve independent Dirichlet problems on subdomains $\Omega^-$,
  $\Omega^+$ using $\lambda^{(k)}$ as the Dirichlet condition for both
  problems at the interface for $\Psi^+$ and $\Psi^-$. This
  corresponds to the subdomain problems such as,
  e.g. \eqref{eq-block}.

\item Now use the discrepancy in the flux between $\Omega^-$ and $\Omega^+$
  to update $\lambda^{(k)}$ by
\ba
\label{eq-lambda}
\lambda^{(k+1)} = \lambda^{(k)} + \theta \jumpi{\eps \nabla \psi^{(k)} \cdot \nu}.
\ea
This step aims at reducing the residual of \eqref{eq-schur}.  It does
not need to be executed, if $\abs{\jumpi{\eps \nabla \psi \cdot \nu}}$ is
smaller than the desired tolerance.
\end{enumerate}
Note that step 1 enforces condition \eqref{eq-jumpd}, and step 2
corrects $\lambda^{(k)}$ based on the degree to which \eqref{eq-jumpz}
fails.

The algorithm we described is a variant on the Neumann-Neumann
algorithm \mpcite{quarteroniV}{Sec. 1.3}, which is an iterative scheme
for the Schur complement system \eqref{eq-schur}, and, in a larger
context, is a preconditioned Richardson iterative scheme. For the
present 1d case the Neumann-Neumann algorithm has a particularly easy
form, since the solution to its Neumann step can be explicitly
found. (In fact, the solution is simply a linear function whose slope
is given on the interface). The algorithm (NN) shown above 
combines all the steps detailed in \mpcite{quarteroniV}{Sec. 1.3}.

The algorithm converges for a suitable $\theta>0$, and the optimal
choices are discussed below.  For
$\lambda^{(0)}$ we can take, e.g., an approximation to $\psi\vert_I$
from a linear guess formed using $\psi_a,\psi_b$ in \eqref{eq-bc0}.

\medskip

{\it Algorithm (NNH)}. The algorithm for the potential equation at
heterojunction, i.e., to solve \eqref{eq-jumppsi}-\eqref{eq-jumpe}, or
in algebraic form, to solve \eqref{eq-dd2}, follows similarly
to Algorithm (NN). It solves for two unknowns
$\Phi_0\mleft,\Phi_0\mright$, but it enforces \eqref{eq-jumppsi} or,
in the discrete form, \eqref{matcalc4}, explicitly.

\medskip

{\it Algorithm (NNC)}. The DDM algorithm for the continuity equation at
a homojunction is similar to Algorithm (NN), since the solution must
satisfy the nonlinear transmission conditions similar to
\eqref{eq-jumpd}-\eqref{eq-jumpz}. However, due to the nonlinearity of
\eqref{eq-nn}-\eqref{eq-np}, or \eqref{eq-nschur}, and to the large
slopes of the solution at the interface, it requires extra care in the
choice of $\theta$. As will be seen in Tables \ref{mat-it},
\ref{mat-it2}, and \ref{delpsi-it}, the choice of relaxation parameter
$\theta$ has a large effect on the performance of the algorithm, and
is greatly influenced by the slope across the interface.

\medskip

{\it Algorithm (NNCH)} for continuity equation with heterojunction
is part of our current work, and will not be described in this paper.

\vspace*{.2cm} {\bf Homojunction examples.}  We first verify that the
algorithm works correctly for this simple case.  The potential
equation is straighforward, and the nonlinear interface equation for
\eqref{eq-nn}-\eqref{eq-np} proceeds simiarly, since we use
iteration-lagging within interface iterations.

\begin{figure}[h!]
  \includegraphics[width=6cm]{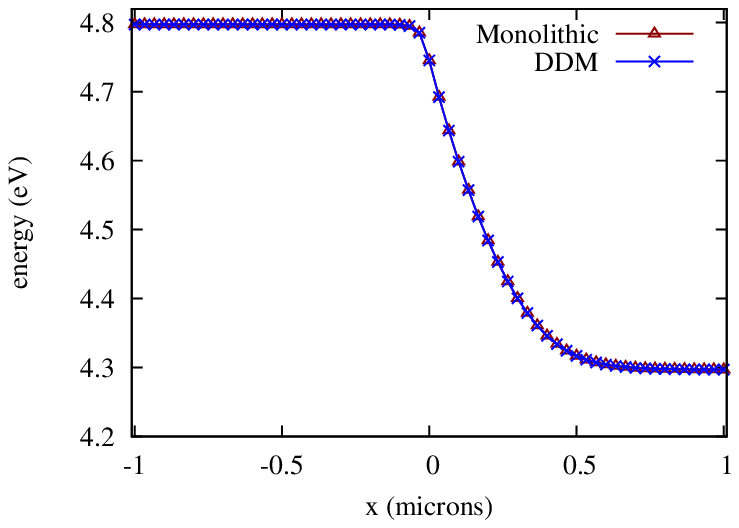}
  \includegraphics[width=6cm]{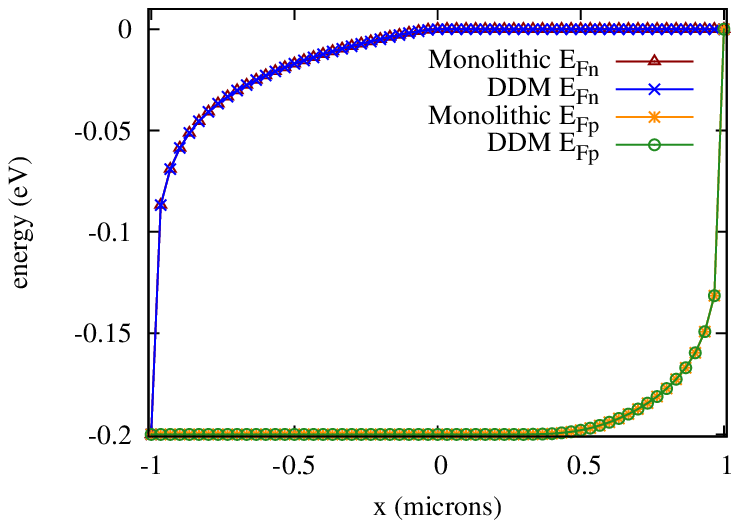}
  \caption{\label{mon-dd-compare}
Comparison of $\Psi,E_{Fn},E_{Fp}$ obtained with the monolithic and DDM solver 
for Silicon n-p homojunction}
\end{figure}

Our experiment is with silicon Si as the material in $\Omega$, with
$\Omega\mleft$ representing a $p-$ doped region and $\Omega\mright$ an $n-$
doped region. 
First we verify that the solver with DDM produces the same results as
that of the monolithic solver, see Figure~\ref{mon-dd-compare}.

Next we discuss the performance of DDM. Table \ref{gummel-ddhomoj}
shows that DDM is mesh-independent for both the potential and the
current-continuity equations, and that the iteration counts $N^I_{it}$
do not vary between Gummel iterations for homojunction case. We note
that the tolerance criteria are currently set the same as in Gummel
iteration, and this could be changed in the future to avoid
oversolving.  We used $\theta=0.25,0.14,0.31$ in Algorithms (NN),
(NNC), (NNC) $\Psi,\Psi_n,\Psi_p$, respectively.

\begin{table}[h!]
  \centering
  \caption{DDM iterations at each Gummel iteration for different $N$
 for Si homojunction}
  \label{gummel-ddhomoj}
  \begin{tabular}{lrrr}
    \hline
    N & $\Psi$ & $\Psi_n$ & $\Psi_p$ \\
    \hline
    1001--4001 & 2 & 5 & 6 \\
    \hline
  \end{tabular}
\end{table}

\begin{figure}[h!]
  \includegraphics[width=8cm]{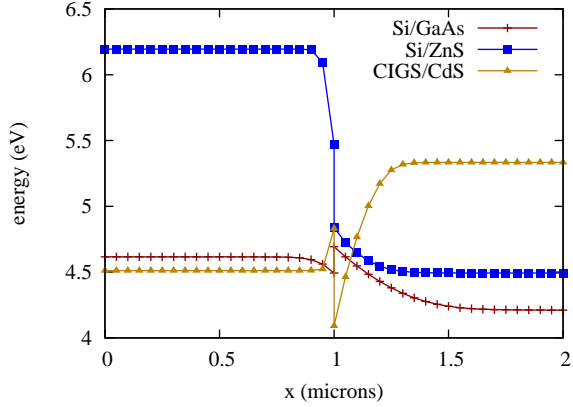}
  \caption{\label{fig-ddpots}Potential across heterojunctions}
\end{figure}

\vspace*{0.2cm} {\bf Heterojunction examples. } We use three examples
of interfaces: \matsig, \matsiz, and \matcigs.  We apply DDM to the
potential equation while the current-continuity equations are solved
monolithically. Even though the potential equation is linear, the
dependence of unknowns on $\mydel \psi$ is nonlinear. We are
interested both in the qualitative results as well as in the
performance of the algorithm as compared to the monolithic solver.

As seen in Figure \ref{fig-ddpots}, the heterojunction causes
discontinuity in the potential, in a way specific to the particular
interface considered. This is consistent with results shown in
Sections~\ref{sec-dfresults}, \ref{sec-cigs}.

Next we discuss various computational aspects of the algorithm.  In
spite of highly varying coefficients and discontinuous solutions, we
verify that DDM-solver and Algorithm (NNH) maintain the second order
of accuracy as expected from FD method. In particular, at every step
of Gummel iteration, its solution, e.g., that to the potential
equation, should exhibit the same convergence order, i.e., second
order, as that for any other self-adjoint elliptic equation, as long
as the solutions are smooth enough. 

To verify, a grid convergence test is performed.  First the solver is
run with a step size of $h = 7.5 \times 10^{-5}$, corresponding to
$N_{max}=40002$ computational nodes.  The solver is then run for
various coarser grids to check for order of convergence of $\psi$ to
the fine grid solution. The error $e(N) \eqdef
\norm{\Psi_N-\Psi_{N_{max}}}{L^2(\Omega)}$ is shown in Table
\ref{fine-grid}.

\begin{table}[h!]
  \centering
  \caption{Grid convergence study, \matsig}
  \label{fine-grid}
  \begin{tabular}{ccc}
    \hline
    $N$	& $e(N)$	 & observed order \\
    \hline
    1002	& $7.81288\times 10^{-8}$	&  \\
    2002	& $1.66832\times 10^{-8}$	& 2.2275 \\
    3002 & $7.05800\times 10^{-9}$	& 2.1216 \\
    4002 & $3.96880\times 10^{-9}$	& 2.0012 \\
    \hline
  \end{tabular}
\end{table}

Further, we check for mesh-independence which holds similarly to that
for homojunction in Table~\ref{gummel-ddhomoj}, see Table~\ref{gummel-dd}.

\begin{table}[h!]
  \centering
  \caption{DDM iterations at each Gummel iteration (GI), \matsig\ heterojunction}
  \label{gummel-dd}
  \begin{tabular}{lrrrrr}
    \hline
    N & GI 1 & GI 2 & GI 3 & GI 4 & GI 5 \\
    \hline
    1002-4002 & 7 & 6 & 4 & 2 & 1 \\
    \hline
  \end{tabular}
\end{table}

The relaxation parameter $\theta$ affects the convergence of Algorithm
(NN), and the optimal choice of $\theta$ is strongly influenced by
material properties. In Table \ref{mat-it} this influence is shown
through comparison of DDM iterations for different materials while
the relaxation parameter is fixed. In particular, a relaxation
parameter good for \matsiz\ has poor performance for \matsig, and
acceptable but sub-optimal performance for \matcigs; these results
correlate with the slope of the potential across the true interface
region $\Omega^I$, see Figure~\ref{fig-ddpots}.  In Table
\ref{mat-it2} we show an optimal $\theta$ for each material determined
by trial and error. In particular, optimal relaxation parameters for
\matcigs\ and \matsiz\ are much closer than those of \matsig.

\begin{table}[h!]
  \centering
  \caption{DDM iterations at each Gummel iteration (GI)
  for various semiconductor materials with $\theta$
  fixed}
  \label{mat-it}
  \begin{tabular}{l|c|c|rrrrrr}
    \hline
    Material & $\mydel \psi$ & $\theta$ & GI 1 & GI 2 & GI 3 & GI 4 & GI 5 & GI 6 \\
    \hline
    \matsiz & -0.63 & 0.005 & 9 & 11 & 10 & 8 & 5 & 2 \\
    \matsig & 0.2 & 0.005 & 225 & 283 & 143 & 21 & 1 & - \\
    \matcigs & -0.74 & 0.005 & 4 & 10 & 7 & 4 & 2 & 1 \\
    \hline
  \end{tabular}
\end{table}

\begin{table}[h!]
  \centering
  \caption{DDM iterations at each Gummel iteration (GI)
  for various semiconductor materials with optimal choice for $\theta$
determined by trial and error}
  \label{mat-it2}
  \begin{tabular}{l|c|rrrrrr}
    \hline
    Material & $\theta$ & GI 1 & GI 2 & GI 3 & GI 4 & GI 5 & GI 6 \\
    \hline
    \matsiz & 0.005 & 9 & 11 & 10 & 8 & 5 & 2 \\
    \matsig & 0.05 & 7 & 6 & 4 & 2 & 1 & - \\
    \matcigs & 0.0051 & 4 & 9 & 6 & 4 & 2 & 1 \\
    \hline
  \end{tabular}
\end{table}

Next, we perform a study of sensitivity of performance of DDM to the
value of $\mydel \psi$. This is important since there may be a large
margin of error in the computed or experimental values of $\mydel
\psi$. We test it for \matsig; see Table \ref{delpsi-it} for behavior
of the DDM for a wide range of $\mydel \psi$. Overall, the method
appears as robust as a Gummel iteration is for a given physical
problem.

\begin{table}[h!]
  \centering
  \caption{Effect of change to $\mydel \psi$ on $N^I_{it}$ at each
    Gummel Iteration (GI), \matsig. $\theta=0.2$}
  \label{delpsi-it}
  \begin{tabular}{lrrrrr}
    \hline
    $\mydel \psi$ & GI 1 & GI 2 & GI 3 & GI 4 & GI 5 \\
    \hline
    0.0 & 8 & 14 & 7 & 1 & 1 \\
    0.1 & 5 & 9 & 6 & 1 & 1 \\
    0.2 & 5 & 6 & 4 & 2 & 1 \\
    0.3 & 7 & 6 & 5 & 2 & 1 \\
    0.4 & 8 & 9 & 6 & 3 & 1 \\
    0.5 & 8 & 7 & 5 & 3 & 1 \\
    \hline
  \end{tabular}
\end{table}

Finally, we compare computational complexity of the monolithic and DDM
solvers. Specifically, we study the number of Newton iterations
$N_{NEWT}$ depending on the discretization $N$ that a monolithic
solver needs with that of subdomain solvers when using DDM. Note that
the complexity of the monolithic solver is $O(N_{NEWT}N)$ per each
Gummel iteration, since we expect that the linear solver involving a
Jacobian in each Newton iteration needs only $O(N)$ operations due to
the three-point stencil of FD discretization. On the other hand, the
subdomain solvers need $O(N_{NEWT}N\mleft)$ or $O(N_{NEWT}N\mright)$
operations, per Gummel iteration, and per each interface iteration $k$
in \eqref{eq-lambda} in Algorithm (NNH). Since $N=N\mleft + N\mright$,
unless the interface iteration \eqref{eq-lambda} converges immediately
or the subdomain solvers are executed in parallel, DDM is about
$N^I_{it}$ times slower in the 1d case than a monolithic solver, which
is well-known.  At the same time, in 2d, for higher order schemes, or
on multicore or multiprocessor environments, the computational
complexity is in favor of DDM, especially for problems close to being
out-of-core.

Still, an important fact is to verify whether $N_{NEWT}$ is mesh
independent for the heterojunction problem, since now the subdomain
solvers may not necessarily react favorably to the progress of
interface iteration.  However, from Table \ref{newt-het} we see that
the monolithic and DDM solvers are both mesh independent, and they
perform similarly in that $N_{NEWT}$ is not much worse for DDM than
that for the monolithic solver. 

\begin{table}[h!]
  \centering
  \caption{Performance of Newton's Method with monolithic and DDM
    solvers for \matsig. Shown are $N_{NEWT}$=maximum number of Newton
    iterations performed in a DDM loop $k$ at each Gummel iteration
    (GI) on subdomains for various mesh sizes}
  \label{newt-het}
  \begin{tabular}{l|l|rrrrr}
    \hline
    & N & GI 1 & GI 2 & GI 3 & GI 4 & GI 5 \\
    \hline
    monolithic& 1002-4002 & 3 & 6 & 3 & 2 & 1 \\
    DDM  & 1002-4002 & 3+3 & 6+5 & 3+5 & 2+5 & 1+5 \\
   \hline
  \end{tabular}
\end{table}

Overall, the performance of DDM is promising for heterojunction, both
in quality of the results and computational complexity. One of many
advantages of DDM is that it can be applied to couple black-box
single-material solvers across interfaces, and these solvers 
can be, in turn, optimized for single materials.

More studies on interface solver are needed, especially as it seems
that it may be oversolving the problem in the first few Gummel
iterations.  Optimal parameters and preconditioners will be needed for
2d geometries.

\section{Discussion and Conclusions}
\label{sec-conclusions}

Numerical simulation of heterojunctions in semiconductor phenomena is
an important tool in material science allowing for design, e.g., solar
cells, and computational implementations in 1d of the basic
heterojunction model have been in use in community and industry codes
such as, e.g., \cite{ampsweb,scapsweb}. In this paper we emphasized
the multiscale character of the problem, focused on its mathematical
structure, and proposed a domain decomposition (DDM) algorithm. Our
applications examples use data from a microscale computational model
based on Density Functional Theory (DFT) which simulates phenomena in
the interface region, and provides the crucial interface data for the
macroscale model. In particular, the DFT model can provide data for
materials that have been predicted but have not yet been
synthesized and/or fully characterized.

The presentation of the coupled bulk subdomain plus heterojunction
model in this paper provides a basis for future analyses, extensions
and improvements. We outlined similarities to other models known from
the literature on fluid flow models in domains with fractures and barriers
or different fluid physics on each of the sides of an interface. There is
substantial current interest in mathematical and computational
literature for those problems, and some results may carry over to the
heterojunction models. In particular, it is clear that an alternative
to using node-centered discretizations, i.e., such as in mixed finite
element methods, needs to be investigated. Mixed finite elements
enforce continuity of fluxes naturally, and do not require doubling of
interface unknowns. In fact, their use was successful for the fluid
flow problems described above, and we plan to investigate their use in
our future work.

The DDM algorithm that we proposed is promising: it decouples the
interface calculations in each step of the Gummel loop from the
subdomain calculations, similarly to how this is done in the Schur
complement formulations. The algorithm makes it possible to use
``black boxes'' or ``monolithic subdomain solvers'' with constant
material parameters, as long as they can use Dirichlet and/or flux
boundary conditions. The algorithm converges easily for the potential
equation, and for continuity equations for homojunctions. The
algorithm for the continuity equation at heterojunctions is in
progress. The DDM formulation can be naturally extended to 2d regions
with complicated geometry of interfaces, or/and to materials with
multiple heterojunctions. In contrast, such extensions may be very
difficult for monolithic solvers in which interface equations are
hard-coded. We plan to address 2d implementation as soon as all he
theoretical and computational issues for the 1d case have been resolved. 

Numerous extensions and refinements of our model are possible. These
include the use of highly refined and adaptive grids based on
a-posteriori finite-element error indicators and/or estimators, as
well as rigorous formulations using other than finite difference
approaches and the use of block-centered grids. Furthermore, for
complicated 2D interfaces, one can consider the use of Immersed
Interface methods (IMM). These have already been proposed for
equations similar to the potential equation with nonhomogeneous jump
in \cite{GongLiLi,HeLinLin,KwakCMAME}, but it is not clear yet how
they may be used for the nonlinear Robin-like condition.


\section*{\normalsize Acknowledgments}
\label{Acknowledgements}

We would like to thank National Science Foundation for supporting this
research via NSF-DMS 1035513 grant "SOLAR: Enhanced Photovoltaic
Efficiency through Heterojunction assisted Impact Ionization".
We also thank Christopher Reidy and Janet Tate for providing
  us with the TEM image in Figure~\ref{fig-TEM}, and Angus Rockett for
  helpful discussions.

We also thank the anonymous referees for their suggestions which
helped to improve the manuscript.

\appendix

\section{Appendix}
\label{sec-appendix}

\subsection{Material and environmental parameters}
\label{sec-df_appendix}

Unless otherwise stated, properties of the materials used in our calculations are shown in Table \ref{tab-matprops}. The values are given in physical (unscaled) units.
Table \ref{tab-ifcprops} gives parameters for particular interfaces or calculations.
The subscripts $n$ and $w$ applied to $N_T$ and $G$ denote the narrow bandgap material and the wide bandgap material respectively.
(By convention, the narrow bandgap material is listed first in the name of the interface; for example ``Si/GaAs'' implies that Si is the narrow bandgap material).
The $\Delta \psi$, $\Delta E_C$, and $\Delta E_V$ quantities are defined by the appropriate property in the wide bandgap material minus this property in the narrow bandgap material.

\begin{table}[H]
\begin{tabular}{l|r|r|r|r|r}
\hline
material & \multicolumn{1}{c}{Si} & \multicolumn{1}{c}{ZnS} & \multicolumn{1}{c}{CIGS} & \multicolumn{1}{c}{CdS} & \multicolumn{1}{c}{GaAs}\\
\hline
$\epsilon$ & 11.9 \cite{Sze} &  8.4 \cite{Sze} & 13.5 \cite{minemoto01the} & 10 \cite{gloeckler03num} & 12.9 \cite{Sze}\\
$\chi$ (eV) &  4.27 (111)  \cite{maierhofer91val} &  3.17 (111) \cite{maierhofer91val} & 4.35 \cite{minemoto01the} & 4.79 \cite{swank67sur} & 4.07 \cite{Sze} \\
$E_g$ (eV) & 1.1 \cite{Sze} &  3.54 \cite{maierhofer91val} & 1.16 \cite{minemoto01the} & 2.42 \cite{wei93ban} & 1.42 \cite{Sze} \\
$N_C$ (cm$^{-3}$) & $2.83\times10^{19}$ \cite{Sze} & $4.3\times10^{18}$ \cite{ruda92book} & $2.2\times10^{18}$ \cite{gloeckler03num} & $2.2\times10^{18}$ \cite{gloeckler03num} & $4.7\times10^{17}$ \cite{Sze}\\
$N_V$ (cm$^{-3}$) & $1\times10^{19}$ \cite{Sze} & $6\times10^{19}$ \cite{ruda92book} & $1.8\times10^{19}$ \cite{gloeckler03num} & $1.8\times10^{19}$ \cite{gloeckler03num} & $7\times10^{18}$ \cite{Sze} \\
$D_n$ (cm$^2$/s) &  37.6 \cite{Sze} & 15.5 \cite{Sze} & 2.6 \cite{gloeckler03num} & 2.6 \cite{gloeckler03num} & 207 \cite{Sze} \\
$D_p$ (cm$^2$/s) &  12.9 \cite{Sze} &  1.0 \cite{seeger10book} & 0.65 \cite{gloeckler03num} & 0.65 \cite{gloeckler03num} & 5.2  \cite{Sze} \\
SRH $\tau_n$ (s)& $1\times10^{-7}$ & $1\times10^{-7}$ & $2\times10^{-9}$ \cite{gloeckler03num} & $1\times10^{-8}$ \cite{gloeckler03num} & $1\times10^{-9}$ \\
SRH $\tau_p$ (s)& $1\times10^{-7}$ & $1\times10^{-7}$ & $1\times10^{-6}$ \cite{gloeckler03num} & $1\times10^{-12}$ \cite{gloeckler03num} & $1\times10^{-9}$ \\
$R_{dc}$ (cm$^3$/s) & $1\times10^{-15}$ & $1\times10^{-10}$ & $1\times10^{-10}$ & $1\times10^{-10}$ & $1\times10^{-10}$ \\
\hline
\end{tabular}
\caption{Parameter values\label{tab-matprops}: materials}
\end{table}

\begin{table}
\begin{tabular}{l|r|r|r|r|}
\hline
interface &               \multicolumn{1}{c}{Si/ZnS} & \multicolumn{1}{c}{CIGS/CdS} & \multicolumn{1}{c}{Si/GaAs} & \multicolumn{1}{c}{Si/Si} \\
\hline
($A_n T^2$) (A cm$^{-2}$) & $3.34\times10^6$ & $2.13\times10^6$ & $7.63\times10^5$ & \\
($A_p T^2$) (A cm$^{-2}$) & $5.86\times10^6$ & $8.67\times10^6$ & $4.62\times10^6$ & \\
$N_{T,n}$ (cm$^{-3}$)     & $5\times10^{15}$  & $-2\times10^{16}$ & $-1\times10^{16}$ & $-5\times10^{17}$ \\
$N_{T,w}$ (cm$^{-3}$)     & $-1\times10^{17}$ &  $2\times10^{17}$ &  $2\times10^{15}$ & $2\times10^{15}$ \\
$G_n$ (cm$^{-3}$s$^{-1}$) &  0, $1\times10^{21}$ &  $1\times10^{21}$ & $1\times10^{21}$  & $1\times10^{21}$ \\
$G_w$ (cm$^{-3}$s$^{-1}$) &  0, $1\times10^{18}$ &  $1\times10^{19}$ & $1\times10^{21}$  & $1\times10^{21}$ \\
$\Delta \psi$ (eV)        & $-0.63$ , $-0.4$ & $-0.74$ & $0.0$ & 0 \\
$\Delta E_C$ (eV)         & $1.73$, $1.5$ & $0.3$ \cite{gloeckler03num} & $-0.02$ & 0 \\
$\Delta E_V$ (eV)        & $-0.71$, $-0.94$ & $-0.96$ & $-0.34$ & 0 \\
\hline
\end{tabular}
\caption{Parameter values\label{tab-ifcprops}: interfaces}
\end{table}

The Shockley-Read-Hall (SRH) recombination lifetimes, $\tau_n$ and $\tau_p$, are highly dependent on dopant density and growth conditions.
In our Si/ZnS simulations we set the SRH lifetimes for Si and ZnS to $1\times10^{-7}$ s.

The SRH lifetime values for the CIGS/CdS calculations are based on 
Ref.~\cite{gloeckler03num}, which gives parameters used by
M.~Gloeckler to establish a baseline model for a CIGS/CdS/ZnO solar
cell.  The parameters reflect alteration from pure material
characteristics both by the presence of high dopant density ($|N_T|
\gg N_i$), as well as by the growth processes used to create layered
thin-film solar cells.

The $R_{dc}$ values are estimated according to whether the material has an electronic structure property known as a direct gap. 
Direct materials are given a value of $1\times10^{-10}$ cm$^3$/s while indirect materials are given a value of $1\times10^{-15}$ cm$^3$/s \cite{Sze}.

$N_V$ values for Si and ZnS are calculated from effective mass data in the cited reference using the convention that the lower energy ``split off'' hole band \cite{Sze} is neglected. (An exception occurs for the Si/GaAs interface, for which we have used the approximation $N_V = 1.8\times 10^{19}$ cm$^3$ for Si, and for which $N_C$ and $N_V$ for GaAs are taken directly from values given in Appendix G of Ref.~\cite{Sze}.)

The values for $D_n$ and $D_p$ given here are typically calculated from data giving the mobilities $\mu_n$ and $\mu_p$, according to
\ba
D_n =  (0.02585 \textrm{ V}) \mu_n (\textrm{cm}^2/(\textrm{Vs})) ,
\ea
with the analogous equation for $D_p$ and $\mu_p$.

For the Si/GaAs calculations, we have used the Si electron affinity value $\chi = 4.05$ eV \cite{Sze}, which is not specific to a particular crystal direction.
In these calculations we have also used $\tau_n = \tau_p = 1\times10^{-9}$ s for both Si and GaAs and neglected direct recombination ($R_{dc}=0$).

The effective Richardson's constants $A_n$ and $A_p$ are estimated \cite{Sze} from $N_C$ and $N_V$ according to
\ba
\label{eq-ant2}
A_n T^2 &= 1.26 \times 10^{-6} (\textrm{A}) \min(N_C\mleft, N_C\mright)^{2/3} \\
\label{eq-apt2}
A_p T^2 &= 1.26 \times 10^{-6} (\textrm{A}) \min(N_V\mleft, N_V\mright)^{2/3} 
\ea
where $T = 300$ K.

The values shown in Table~\ref{tab-ifcprops} for the Si/ZnS interface indicate both thermal equilibrium and illuminated ($G > 0$) calculations, while the offsets shown correspond to experiment \cite{maierhofer91val} (first value) and the DFT calculations (second value). 

In all simulations, the recombination velocities defined in Section~\ref{sec-bc} are given values between $1\times10^6$ and $1\times10^7$ cm/s, as these velocities are typically less than or approximately equal to the average (thermal) electron velocity of about $1\times10^7$ cm/s.


\subsection{Scaling for the continuum model}
\label{sec-appendix-scale}

The dimensionless quantities using in the model formulated in Section
\ref{sec-model} must generally be multiplied by appropriately
dimensioned scaling constants.  Our scaling in many respects follows
that of \cite{Selber}.

For charge densities such as $n$, $p$, $N_C$, and $N_T$, it is
conventional to use as physical units the number of individual
carriers or atoms per cubic centimeter, or simply cm$^{-3}$.  It is
thus convenient to choose a scaling constant $C_0$ such as
\ba
C_0 = \max_{\Omega} |N_T (\textrm{cm}^{-3})|,
\ea
or to use scaling based on $N_i$ or $N_C$ and $N_V$.  This simple
definition of the dimensionless quantities $n$, etc.~will lead to a
formal modification of the Poisson equation as given in
\eqref{eq-psi}.  (Equation \eqref{eq-psi} omits constants for
simplicity but requires a less intuitive scaling for the charge
densities.)

The potential variables such as $\psi$ and $\psi_n$ can play a dual
role as both scaled electrostatic potential quantities and as scaled
energies of single charge carriers.  Properly, $\psi$ is an
electrostatic potential, originally in Volts (V), scaled by the
thermal energy $k_B T$ divided by the electronic charge $e$ (potential
= energy / charge).  $k_B \approx 8.61733\times10^{-5}$ eV/K is the
Boltzmann constant and $T$ is the absolute temperature in Kelvins.  In
calculations we use $T = 300$ K to approximate room temperature.  The
electronic charge $e$ is the positive elementary unit of charge; the
charge of an electron is $-e$.  The electron Volt, eV, is an energy
equal to the electronic charge multiplied by 1 Volt. Physically, an
electron (hole) in a region of potential 5 V has electrostatic energy
$-5$ eV ($+5$ eV).  We define the scaling constant
\ba
V_0 &= k_B T / e \notag \\
&= 8.61733\times10^{-5} \textrm{ eV/K} \times 300 \textrm{ K} / e \notag \\
&= 0.0258520 \textrm{ V}
\ea
The dimensionless $\psi$ can be multiplied by $V_0$ to obtain the
physical voltage, and a change in applied voltage of $+0.1$ V at
boundary $a$ corresponds to a change of $+0.1$ eV / $V_0$ in the
dimensionless quantity $\psi_a$.  To obtain the energies of electrons
and holes in the conduction and valence bands, we multiply physical
potential quantities such as $V_0 \psi$ by $\pm e$.  When these
energies are expressed in eV, the multiplication above does not change
the numerical value, and thus the value of $V_0$ can also be thought
of as the value of the energy scale $k_B T = V_0 e$.

Using the scaling above, the electrostatic Poisson equation
represented by \eqref{eq-psi} requires a factor involving the physical
constants $\epsilon_0 = 8.85419 \times 10^{-14}$ C/(V cm), the
permittivity of free space, and the electronic charge $e =
1.602177\times 10^{-19}$ C, where C denotes Coulomb, the SI unit of
charge.  The Poisson equation for the scaling described above is thus
\ba
-V_0 \eps_0 / (e C_0 X_0^2) \nabla \cdot (\eps \nabla \psi) = p - n + N_T,
\ea
where the variables $\psi$, $n$, $p$, and $N_T$ are dimensionless, $e$ is expressed in Coulombs, and $X_0$ is an optional distance scaling factor in cm. We have used
\ba
X_0 = 10^{-4} \textrm{ cm}.
\ea

We use a scaling factor $D_0$ to make $D_n$ and $D_p$ dimensionless,
\ba
D_0 = \max_{\Omega} (D_n, D_p ({\textrm{cm}^2/\textrm{s}})).
\ea
The result of the scaling of $D_{n(p)}$ and $x$ is that time is
scaled. For example the SRH recombination constants $\tau_{n(p)}$ are
scaled by \ba t_0 = X_0^2/D_0. \notag \ea The dimensionless mobilities
$\mu_{n(p)}$ have been scaled by $D_0/V_0$.

An important quantity is the physical current density which is given in A/cm$^2 = \textrm{C/(s cm}^2$).
The resulting scaling factor for current densities is
\ba
J_0 = e \frac{D_0 C_0}{X_0},
\ea
where $e$ is in Coulombs.

Scaling of the remaining quantities used in the continuum model can be
derived in terms of the above scaling constants by inspection of the
equations that use and/or define them.

\subsection{Equivalence of transmission conditions to the algebraic
DDM formulation via second order finite differences}
\label{sec-appendix-dd}

Here we show that with proper discretization of interface condition
expressing continuity of fluxes as in the transmission condition
\eqref{eq-jumpz} we obtain the algebraic system \eqref{eq-dd2}. The
argument presented below also works for the heterojunction system
\eqref{eq-ddh}.

First, assume that the material constants $\eps$ are not the same on
both sides of the interface which accounts later for the case of
heterojunction.

We note that a one-sided approximation $\frac{\partial \psi
}{\partial x}\vert\ileft \approx \frac{\Psi\zleft-\Psi_{-1}}{h}$ at
the interface $I$ is only first order accurate, and a similar
approximation $\frac{\partial \psi }{\partial x}\vert\iright \approx
\frac{\Psi_{1}-\Psi\zright}{h}$ together with \eqref{eq-jumpz} result
in
\ba
\label{eq-j0}
\eps\mleft \Psi\zleft + \eps\mright \Psi\zright 
-\eps\mleft \Psi_{-1}-\eps\mright \Psi_{1}=0,
\ea
thus do not allow any charges at $I$, and lead to an
inaccurate solution. 

A second order approximation $ \frac{\partial \psi }{\partial
  x}\vert\ileft \approx \frac{\Psi^*_{1}-\Psi_{-1}}{2h}$ introduces an
additional ``ghost'' unknown $\Psi^*_{1}$, which can be eliminated
using a discretization of \eqref{eq-psi} that would be posed at
$x\zleft$, i.e., \eqref{eq-fd} for $j=0\mleft$
\ba
\label{eq-j1}
2 \eps\mleft \Psi\zleft -\eps\mleft \Psi_{-1}-\eps\mleft \Psi^*_{1}=Q_0\mleft,
\ea
Similarly, for an approximation to $ \frac{\partial \psi }{\partial
  x}\vert\iright = \approx \frac{\Psi_{1}-\Psi^*_{-1}}{h}$ we have
\ba
\label{eq-j2}
2 \eps\mright \Psi\zright -\eps\mright \Psi^*_{-1}-\eps\mright \Psi_{1}=Q_0\mright,
\ea
From \eqref{eq-jumpz} we have, using these approximations, that
\begin{multline*}
\eps\mleft \frac{\partial \psi }{\partial
  x}\vert\ileft 
\approx
\eps\mleft \frac{\Psi^*_{1}-\Psi_{-1}}{h}\\=\eps\mright \frac{\Psi_{1}-\Psi^*_{-1}}{h}
\approx \eps\mright \frac{\partial \psi }{\partial
  x}\vert\iright 
\end{multline*}
and multiplying by $h$ we get
\ba
\label{eq-j3}
\eps\mleft (\Psi^*_{1}-\Psi_{-1})- \eps\mright (\Psi_{1}-\Psi^*_{-1}) =0
\ea
Adding \eqref{eq-j1}-\eqref{eq-j3}, and dividing by $2$ eliminates the
ghost unknowns $\Psi^*_{1}, \Psi^*_{-1}$, and we obtain
\ba
\label{eq-jall}
\eps\mleft \Psi\zleft + \eps\mright \Psi\zright
-\eps\mleft \Psi_{-1}-\eps\mright \Psi_{1}=\frac{Q_0\mleft+Q_0\mright}{2}
\ea
which is a proper discretization of \eqref{eq-average}.

{\bf Homojunction case.} If $\eps\mleft=\eps\mright=\eps$ we obtain
immediately that \eqref{eq-jall} is the same as the second to last row
of \eqref{eq-dd2}, with $A_{0,0} = 2 \eps$ as in \eqref{eq-dd}. 

{\bf Heterojunction case.} In the case $\eps\mleft \neq \eps\mright$
we see that \eqref{eq-jall} is the same as the second to last row of
\eqref{eq-ddh}, with $A_{0,0}\mleft = \eps\mleft$ and $A_{0,0}\mright
= \eps\mright$.

\subsection{Schur complement form for heterojunction}
\label{sec-appendix-ddh}

Here we show details leading to \eqref{Xi-heterdd},
\eqref{Phi-heterdd}.  We rewrite \eqref{eq-ddh} to see that it gives the
four equations
\ba
\label{matcalc1}
A_{-,-} \Psi^- + A_{-,0} \Psi^-_0 &=& Q^-,
\\
\label{matcalc2}
A_{+,+} \Psi^+ + A_{+,0} \Psi^+_0&=& Q^+,
\\
\nonumber
A_{0,-} \Psi^- + A_{0,+} \Psi^+ 
+ A_{0,0}^- \Psi^-_0 \\+ A_{0,0}^+ \Psi^+_0 
\label{matcalc3}
&=&\avei{Q^I},
\\
\label{matcalc4}
\Psi_0^+ - \Psi_0^- &=& \mydel \psi.
\ea
Using \eqref{matcalc4} in \eqref{matcalc2}-\eqref{matcalc3} we obtain:
\ba
\label{matcalc6}
A_{+,+} \Psi^+ + A_{+,0}(\Psi_0^- + \mydel \psi) &=& Q^+,
\\
\nonumber
A_{0,-} \Psi^- + A_{0,+} \Psi^+ + A_{0,0}^- \Psi^-_0
\\
+ A_{0,0}^+(\Psi_0^- + \mydel \Psi) &=& \lbrace Q^I \rbrace.
\label{matcalc7}
\ea
Now we use these and \eqref{matcalc1} to solve for $\Psi^-$ and $\Psi^+$,
\ba
\label{matcalc8}
\Psi^- = A_{-,-}^{-1} Q^- - A^{-1}_{-,-} A_{-,0} \Psi_0^-,
\\
\label{matcalc9}
\Psi^+ = A_{+,+}^{-1} Q^+ - A_{+,+}^{-1} A_{+,0} \Psi^-_0 - A_{+,+}^{-1} \mydel\psi.
\ea
Substituting \eqref{matcalc8} and \eqref{matcalc9} into \eqref{matcalc7} then gives us
$\Xi \Psi_0^- = \Phi$ with $\Xi$ given by \eqref{Xi-heterdd} and $\Phi$ given by 
\eqref{Phi-heterdd}.

If $\mydel \psi=0$, then the Schur-complement form \eqref{eq-schur-ddh}
reduces to \eqref{eq-schur}. 


\bibliographystyle{acs}
\def\cprime{$'$}


\end{document}